\documentstyle[a4,11pt]{article}
\input{epsf}

\newcommand{\slL}{\raise.15ex\hbox{$/$}\kern-.53em\hbox{$L$}}
\newcommand{\slP}{\raise.15ex\hbox{$/$}\kern-.53em\hbox{$P$}}
\newcommand{\slR}{\raise.15ex\hbox{$/$}\kern-.53em\hbox{$R$}}
\newcommand{\slQ}{\raise.15ex\hbox{$/$}\kern-.53em\hbox{$Q$}}
\newcommand{\slK}{\raise.15ex\hbox{$/$}\kern-.53em\hbox{$K$}}

\newcommand{\be}{\begin{equation}}
\newcommand{\ee}{\end{equation}}     
\newcommand{\bea}{\begin{eqnarray}}
\newcommand{\ena}{\end{eqnarray}}

\def\build#1\over#2{\mathrel{\mathop{\kern 0pt#2}\limits_{#1}}}

\font\cmr=cmr7

\font\tenimbf=cmmib10 at 12pt
\font\sevenimbf=cmmib10 at 7pt
\font\fiveimbf=cmmib10 at 5pt
\newfam\imbf
\textfont\imbf=\tenimbf
\scriptfont\imbf=\sevenimbf
\scriptscriptfont\imbf=\fiveimbf
\def\imb{\fam\imbf\tenimbf}

\begin{document}

\def\thefootnote{\alph{footnote}}

\begin{titlepage}
\def\thefootnote{\alph{footnote}}
\title{\bf{Breakdown of the Hard Thermal Loop}\\
       \bf{ expansion near the light--cone}}
\author{\def\thefootnote{\alph{footnote}}
P.~Aurenche$^{1,2,}$\footnotemark\ %
$^{,}$\footnotemark\ ,
 F.~Gelis$^{2,}$\footnotemark\ , 
R.~Kobes$^{3,}$\footnotemark\ ,
 E.~Petitgirard$^{3,}$\footnotemark}
\def\thefootnote{\alph{footnote}}
\maketitle
\begin{enumerate}
\item CFIF, Instituto Superior T\'ecnico, Edificio Ci\^enca %
(f\'\i sica),\\
P-1096 Lisboa, Codex, Portugal
\item Laboratoire de Physique Th\'eorique ENSLAPP,\\
B.P. 110, F-74941 Annecy-le-Vieux Cedex, France
\item Physics Department and Winnipeg Institute
for Theoretical Physics,\\
University of Winnipeg,
Winnipeg, Manitoba R3B 2E9, Canada
\end{enumerate}
\vskip -1cm
\begin{abstract}
We discuss the bremsstrahlung production of
soft real and virtual photons in a quark-gluon plasma at thermal
equilibrium beyond the Hard Thermal Loop (HTL) resummation.
The physics is controlled by the ratio $Q^2/q_0^2$ of the
virtuality to the energy.  When $Q^2/q_0^2\ll g^2$, where
$g$ is the strong coupling constant, the emission rate
is enhanced by a factor $1/g^2$ over the HTL results due to light-cone
singularities and the bremsstrahlung is induced by scattering
of the quark via both transverse and longitudinal soft gluon exchanges.
When  $Q^2/q_0^2$ increases, the enhancement factor is given 
by $q_0^2/Q^2$. When this ratio is near unity, the bremsstrahlung
contribution is of the same order as the rate predicted
by the HTL resummation. In that case, the bremsstrahlung
is induced by both soft and hard gluon exchanges.
\end{abstract}
\def\thefootnote{\alph{footnote}}
\footnotetext[1]{On leave of absence 
from ENSLAPP, B.P. 110, 
F--74941 Annecy-le-Vieux Cedex, France}
\footnotetext[2]{aurenche@lapp.in2p3.fr,\ \ \ \ \ \ $^c$gelis@lapp.in2p3.fr}
\footnotetext[4]{randy@theory.uwinnipeg.ca,\ \ \ \ \ \ $^e$petitg@theory.uwinnipeg.ca}

\vskip 4mm
\centerline{\hfill ENSLAPP--A--614/96,  WIN--96--13,   hep-ph/9609256\hglue 2cm}
\vfill
\thispagestyle{empty}
\end{titlepage}

\section{Introduction}

Several methods have recently been used for calculating the rate of
soft real or virtual photons in a hot quark-gluon plasma.  These
methods lead to drastically different physical pictures and orders of
magnitude for the calculated rates.  In the following we will always
assume the plasma to be in thermal equilibrium and the QCD coupling
constant $g \ll 1$.  We furthermore assume that the produced
photon does not interact with the quark-gluon plasma due to the smallness
of the electric charge $e$ compared to the strong interaction coupling:
in other words the photon is not thermalized.
\par
In the hard thermal loop (HTL) resummation
analysis \cite{rp,brat,tay} the photon is radiated off in processes which
always involve an interaction between the plasma particles which are
dominated by soft (energy $\sim gT$) quasi-fermion exchanges
\cite{yuan}.  In the case of real photon emission, a strict
application of the HTL rules leads to a prediction with a logarithmic
singularity associated to collinear divergences \cite{schiff,pat}. 
\par
In a semi-classical approach \cite{cleymans,cleyman1,goloviz}, 
the photon is radiated
off by a hard quark (energy $\sim T$) interacting with quarks and
gluons in the plasma via the exchange of soft gluons, the interactions
being assumed to be screened by the Debye mass.  Multiple
re-interactions of the radiating quark in the plasma regularize
potential collinear singularities in the lowest order process and,
more importantly, suppress the photon spectrum at small frequencies
due to the Landau-Pomeranchuk-Migdal (LPM) effect \cite{lanpom}. 
For a different approach of this effect, see \cite{quack,knoll}.
\par
The regularization of divergences in the bremsstrahlung process has
also been achieved, under somewhat different hypotheses, by the resummation 
of soft photon emission with a single interaction of the radiating 
quark in the plasma \cite{weldon,gupta}.  We will not consider this 
possibility as it affects the photon spectrum at very small
momenta, of order $eT$, while we are interested here in momenta
of order $gT$.

In the following, we come back to the problem of real or virtual soft photon
emission, in the framework of thermal field theory, showing the necessity of
going beyond the HTL approximation for obtaining consistent predictions.  The
picture which emerges is rather different from the above quoted results.
Compared to the HTL approach we find, in fact, that there is no contribution
from hard thermal loops at the expected leading order in $g$.  However some non
leading terms in the HTL approximation become enhanced by singularities near
the light-cone and dominate over the would-be HTL order \cite{letter}.  In the case of real
or quasi-real photons (virtualities $Q^2/q_0^2 \ll 1$) the dominant
diagrams are the bremsstrahlung ones in agreement with the hypotheses of
\cite{cleymans,cleyman1,goloviz} and we find that, for 
$Q^2/q_0^2 \le g^2$, the production rate is $R\sim e^2 g^2 T^4/q_0^2$ 
modulo some logarithmic factor.  The results are shown to be
independent of the covariant gauge parameter.  Unlike in
\cite{cleymans,cleyman1,goloviz} we treat the interaction between quarks and
gluons in the plasma ``exactly", in the framework of thermal field theory, and
we do not find that the static approximation is a good one as exchanges of
``magnetic'' gluons are as important as that of ``electric'' ones. 

 In summary, the result of our analysis concerns photons of 
momentum much less than $T$ radiated by hard quarks in a plasma:
for virtualities $Q^2/q_0^2 \ll 1$ we find an enhancement, due
to light-cone singularities, over the rate $R \sim e^2 g^4 T^4/q_0^2$ (modulo 
logarithmic factors) expected in the HTL approximation;
for virtualities $Q^2/q_0^2 \sim 1$ no enhancement occurs but the bremsstrahlung 
contributions are of the same order as the ``soft-fermion'' contributions 
calculated in \cite{yuan}. This extends out of the light--cone
our previous results
on photon production on the light--cone \cite{letter}.

Problems in the HTL expansion associated to light-cone singularities have 
already been discussed several times in the literature: as mentioned above
they render the rate of soft real photon production logarithmically
divergent \cite{schiff,pat,niega}. In fact, already in their original
paper \cite{brat} Braaten and Pisarski discuss mass divergences,
$i.e.$ terms such as $g^2 T / (\omega - p)$ with $\omega \rightarrow p$,
and they mention that these terms could spoil the HTL expansion. 
In order to solve this problem it has been proposed recently
\cite{rebhan,rebha1,rebha2}
to extend the HTL action to include thermal corrections (essentially 
the effective thermal mass) on hard propagators but no general proof exists
that this improvement is sufficient or complete. A particularly
interesting feature of the problem we consider here
 is that the mass singularities change the expected
order of the leading HTL prediction and enhance it by a factor which
may be as large as $1 / g^2$ in some cases.

\section{Discussion of the existing approaches}

This part is devoted to a critical appraisal of the previous
calculations of the real and virtual photon production rates in a 
quark-gluon plasma.

\subsection{Weaknesses of the Hard Thermal Loop expansion}

One of the first applications
 of the HTL resummation scheme was the calculation
of the production rate of a soft photon, of virtuality $Q^2$, at rest in a 
quark-gluon plasma \cite{yuan}. Before resummation is taken into account, the 
photon is produced by the annihilation of a soft quark-antiquark pair in
the plasma. The imaginary part of the vacuum polarization diagram,
which is proportional to the invariant
 production rate $R$ up to a statistical 
weight, is found
to be 
\bea 
\mbox{Im} \Pi^\mu\,_\mu (Q) \sim e^2 Q^2 (1 - 2 n_{_F}({Q \over 2}) )
\sim e^2 g^3 T^2 \qquad\hbox{\rm for}\quad q,q_0 \sim gT
\label{eq:born}
\ena
In the HTL approach the diagrams to be calculated are shown in  
Fig.~\ref{figsoft}. In fact, only the first one, with effective quark
propagators and effective $q-{\overline q}-\gamma^*$ vertices, is said to
contribute: the self-energy diagram with the gluon loop in fact does not
exist because there is no $g-g-\gamma$ effective vertex
(Furry's theorem) and the two tadpole diagrams vanish 
because the traces
of the corresponding 4-point effective vertices also vanish. 
A compact
expression, involving convolutions of the effective soft 
fermion spectral
functions has been obtained by Braaten, Pisarski and 
Yuan \cite{yuan}
and one can estimate the order of magnitude of their result 
to be as in
Eq.~(\ref{eq:born}). An important fact concerns the statistical factor
$(1 - 2 n_{_F}({Q \over 2}))$ associated to the annihilating $q$ and 
$\overline q$ which becomes 
$(1 - n_{_F}(\omega) - n_{_F}({Q \over 2}-\omega))$, where $\omega$
is the energy of one of the annihilating soft fermion in the 
HTL approach:
for soft energies it brings an extra suppression factor $g$ 
to the photon
production rate.
\par
Consider now, for example, the last diagram with the gluon tadpole
and in particular the order
of $\Pi^{00}$ or $\Pi^{ij}$. Using the HTL rules, the
$g-g-\gamma-\gamma$ effective vertex is found to be of order
$e^2$ while
the soft effective gluon loop yields \cite{brat}
\be  
\int_{{\mbox {\scriptsize soft}}\ k} d^3 k {n_{_B}(k) \over k} 
\sim g T^2
\ee
because of the enhancement due to the Bose-Einstein statistical factor
$n_{_B}(k)$. We then find
\be 
\Pi^{00} \sim \Pi^{ij} \sim e^2 g T^2.
\ee
Of course, we have a cancelation when taking the trace $\Pi^\mu{}_\mu$:
this is why this diagram is said to be vanishing in the HTL framework.
The important point is that the vanishing of the HTL contribution
occurs in this diagram at the order $e^2 g T^2$ while the calculated
 soft fermion loop contribution
is of order $e^2 g^3 T^2$. Therefore, non-leading terms (in the HTL 
sense) in the gluon tadpole may be as large as the soft fermion
loop contribution. 
This is indeed the case and these non-leading terms will play a very
important role. A similar discussion may be given concerning the 
fermion
tadpole Fig.~\ref{figsoft}-(c) which should however be suppressed 
by a power of
$g$ compared to Fig.~\ref{figsoft}-(d) because of the lack of 
Bose-Einstein
enhancement at soft momenta.

In terms of physical processes Fig.~\ref{figsoft}-(a) involves 
the first three amplitudes
shown in Fig.~\ref{figampl} (and their crossing symmetric ones) where
the photon is radiated off either by a soft quasi-fermion (a) or by
a hard on shell fermion (b and c): the imaginary part of 
Fig.~\ref{figsoft}-(a)
is in fact constructed from squaring Fig.~\ref{figampl}-(a) or 
taking
the interference of (a),(b) and (c). 
On the other hand ``squaring" the diagram Fig.~\ref{figampl}-(b)
or (c) yields the soft fermion tadpole diagram of Fig.~\ref{figsoft}-(c).
In all cases though,
the interactions between hard particles in the plasma are mediated by
soft fermion exchanges. The absence of processes where a gluon
is exchanged between the partons in the plasma, which are intuitively
expected to be relevant, seems to indicate that the calculation based
on the soft fermion loop is not complete. The expected amplitudes are, in fact,
contained in Fig.~\ref{figsoft}-(d) and are obtained when ``cutting"
 through
the effective gluon and the effective 4-point function: these are shown
in Fig.~\ref{figampl}-(d) and (e). In this set of diagrams
the hard gluon line can be replaced by a hard quark line,
since the radiating quark can be scattered by any kind of parton in
the plasma. The physical
process involved is the bremsstrahlung of the photon by a hard quark
which scatters in the plasma via the exchange of a soft gluon. The calculation
of the diagram of Fig.~\ref{figsoft}-(d), in thermal field theory, will be 
the main purpose of this paper and it will be found that indeed its 
contribution to ${\rm Im}\,\Pi^\mu{}_\mu$
is at least of order $e^2 g^4 T^3/q_0$ and that, in the case
of quasi-real photons, of virtuality $Q^2/q_0^2 \ll 1$, it
is further enhanced by powers of $1 / g$.

To be complete, one should also mention that the soft fermion loop
contribution to real photon production has been calculated 
\cite{schiff,pat} and the result is as in Eq.~(\ref{eq:born}) with the
added feature of logarithmic mass singularities due to the emission
of the photon collinear to the hard massless quark. These singularities
are regularized by giving the quark its asymptotic thermal mass \cite{niega}.

\subsection{Semi--classical approach: LPM effect}

The diagrams of Fig.~\ref{figampl}-(d) and (e) have been considered in a series of papers
by Cleymans, Goloviznin and Redlich \cite{cleymans,cleyman1,goloviz}.  In fact,
the purpose of these works was to study the effect of multiple rescattering of
the radiating quark in the plasma:  they find that even when working with
massless quarks the rate of production of on-shell photons is finite due to
rescattering effects, and, more importantly, they show that the shape of the
photon spectrum is modified, at small energies, and becomes $1 / \sqrt{q_0}$
rather than the usual logarithmically divergent bremsstrahlung spectrum in 
$1 / q_0$:  this is an illustration of the Landau-Pomeranchuk-Migdal effect
\cite{lanpom}.  The results of \cite{cleymans,cleyman1,goloviz} are obtained in
the semi-classical approximation (the quark follows a classical path) and it is
assumed that the interaction of the quark in the plasma is as at $T=0$, the only
thermal input being the Debye mass, introduced by hand to regularize the forward
singularity of the quark scattering amplitude.  With the LPM effect taken into
account their result yields ${\rm Im}\,\Pi^\mu{}_\mu\sim e^2 g^2 T^2
(T/q_0)^{1/2}$,
whereas without the LPM effect, the case we concentrate upon below,
 one obtains
${\rm Im}\,\Pi^\mu{}_\mu \sim e^2 g^2T^3/q_0$. 
 We will come back to a detailed comparison of this result
with our calculation after our results have been presented.


\vskip 1cm\goodbreak

\section{Generalities of the calculation}

This section is devoted to explain the notations and to give general
results to be used later whatever the virtuality of the emitted photon is.

\subsection{Some notations and preliminaries}

The invariant rate of production, per unit time and per unit volume of the plasma,
 of a real photon is given to first order 
in $\alpha$ ($i.e.$ this formula takes only into account the emission
by a quark of a single photon, but does not consider the
possibility of emitting more than one photon) by \cite{gale}:
\begin{equation}
  \label{rate}
 R\equiv q_0 \frac{d N}{d^3 q d^4x} = - \frac{1 }{ (2 \pi)^3} n_{_{B}} (q_0)
  \ \hbox{\rm Im}\, \Pi^\mu\,_\mu (Q).
\end{equation}
For a lepton pair of invariant mass squared $Q^2$ the rate, 
integrated over 
the leptons kinematical variables, is 
\be
\frac{d N}{d q_0 d^3 q d^4x} = - \frac{\alpha }{ 12 \pi^3} 
\frac{1}{Q^2} n_{_{B}} (q_0)
  \ \hbox{\rm Im}\, \Pi^\mu\,_\mu (Q).
\label{ratevir}
\end{equation}
It is worth noticing that these formulae, although giving only the
dominant term in $\alpha$ of the quark, are exact to all orders in
$\alpha_{_{S}}$.  We will need to compute the imaginary part
of the photon self-energy, which is given at lowest order 
in $\alpha_{_{S}}$ by the two loop
diagrams of Fig.~\ref{fighard}, to which one should
also add the diagram with the
self-energy correction on the upper fermion line (not represented).
These diagrams are an explicit representation of the gluon tadpole of
Fig.~\ref{figsoft}-(d) and, as explained before, they will be calculated
beyond the HTL approximation. In the following, a summation over 
the transverse (T), longitudinal (L) and gauge dependent (G)
modes in the gluon propagator is always assumed.

The notations for the momenta are explained in Fig.~\ref{fighard}.
  In the
kinematical domain we consider, the fermion in the loop has a hard
momentum, which means that the components of its momentum are of order
$T$, and the exchanged gluon as well as the external photon have
 momentum components
 much smaller than $T$.  


\subsubsection{The ``retarded-advanced" formalism}

In order to compute the needed imaginary part, we will proceed by using
the ``retarded-advanced" formalism \cite{ra,van,van2}, which is one of
the simplest versions of the real-time formalisms.  This formalism is
obtained from the more common ``1-2" formalism
\cite{fett,niem,kobe,land} by a change of basis. The result 
of such a
transformation is a formalism with a new $2\times 2$ matrix
propagator and new vertices. A nice feature of the
``retarded-advanced" formalism lies in the fact that it takes into
account, in an optimal way, the Kubo-Martin-Schwinger identities valid at
thermal equilibrium, so that among the four components of the matrix
propagator only two are non zero; more precisely, the matrix propagator
we get for a real scalar field is (for fermions or gauge fields, one has
to take into account also the Dirac or Lorentz structure):
\begin{equation}
\Delta(P)=\pmatrix{
&\Delta_{_{R}}(P)&0\cr
&0&\Delta_{_{A}}(P)\cr
},
\end{equation}
where the retarded and advanced 
propagators are given by
\begin{equation}
\Delta_{_{R,A}}(P)\equiv{{i}\over{P^2-M^2\pm i \epsilon p_0}},
\ \ \ \epsilon > 0.
\end{equation}
Another feature of this matrix propagator is that it no longer
contains any information relative to the thermal equilibrium. Of course,
as said before, the change of basis for the
propagator has as a consequence a change
in the vertices to be used, and the price to pay for these
simple propagators is somewhat intricate vertices. For example, for a
$QED$-like vertex with entering momenta $P,Q$ and $-R$, $Q$ being
the photon momentum, we have:
\begin{eqnarray}
&&g_{_{AAA}}(P,Q,-R)=g_{_{RRR}}(P,Q,-R)=0,\nonumber\\
&&g_{_{RRA}} (P,Q,-R) = g_{_{ARR}} (P,Q,-R) =g_{_{RAR}} (P,Q,-R) =e,
  \nonumber\\
&&g_{_{RAA}}(P,Q,-R) = - e\ (1+n_{_{B}}(q_0)-n_{_{F}}(r_0)),\nonumber\\
&&g_{_{ARA}} (P,Q,-R) = -e\ (1-n_{_{F}}(p_0)-n_{_{F}}(r_0)),
\end{eqnarray}
where $n_{_{B}}(q_0)\equiv1/\left(e^{\beta q_0}-1\right)$,
$n_{_{F}}(p_0)\equiv1/\left(e^{\beta p_0}+1\right)$,
and $\beta\equiv1/T$ is the inverse temperature. Therefore, in this
formalism, all the thermal information is put into the vertices.
The thermal content of an $n$--point function is represented by
the external $R/A$ indices. Inside a loop a sum is made over
the internal thermal indices with
appropriate internal propagators joining them.

\subsubsection{Effective soft gluon propagator}
Since the exchanged gluons can be soft in our diagrams, we will
need to use effective propagators taking into account
the Hard Thermal Loop corrections \cite{brat,tay}. 
We choose to use a covariant
gauge in which the effective gluon propagator 
is decomposed into its transverse, longitudinal and gauge components:
\begin {equation}
-D^{\mu\nu}(L)\equiv{P^{\mu\nu}_{_{T}}(L)}\Delta^{^{T}}(L)
+{P^{\mu\nu}_{_{L}}(L)}\Delta^{^{L}}(L)
+\xi\,{P^{\mu\nu}_{_{G}}(L)}\Delta^{^{G}}(L),
\label{gluon}
\end{equation}
where $\xi$ is the gauge parameter, and
with
\begin{eqnarray}
&&\Delta^{^{T,L}}(L)\equiv{i\over{L^2-\Pi_{_{T,L}}(L)
}}
\nonumber\\
&&\Delta^{^{G}}(L)\equiv{i\over{L^2}
}.
\end{eqnarray}
The appropriate choice of $l_0 \pm i \epsilon$ is implicitly 
understood 
when considering R, A propagators.
 The tensors $P^{\mu\nu}_{_{T,L,G}}(L)$ are the
transverse, longitudinal and gauge projectors whose explicit 
expression are \cite{land,wel,klim,pisars}     
\begin{eqnarray}
&&P^{\mu\nu}_{_{T}}(L)=\gamma^{\mu\nu}-{{\kappa^\mu\kappa^\nu}
\over{\kappa^2}}\nonumber\\
&&P^{\mu\nu}_{_{L}}(L)=U^\mu U^\nu+{{\kappa^\mu\kappa^\nu}
\over{\kappa^2}}-{{L^\mu L^\nu}\over{L^2}}\nonumber\\
&&P^{\mu\nu}_{_{G}}(L)={{L^\mu L^\nu}\over{L^2}}
\end{eqnarray}
where $\gamma^{\mu\nu}\equiv g^{\mu\nu}-U^\mu U^\nu$, $\kappa^\mu
\equiv \gamma^{\mu\nu}L_\nu$, and where $U$ is the mean velocity of
the plasma in the current frame. As is usual, in the following we
will work in the rest-frame of the plasma in order to have 
simpler expressions.

The explicit form of the
functions $\Pi_{_{T,L}}(L)$ are\cite{wel,klim}
\begin{eqnarray}
&&\Pi_{_{T}}(l,l_0)=3 m^2_{\hbox{\cmr g}}\left[
{{x^2}\over 2}+{{x(1-x^2)}\over{4}}\ln\left({{x+1}\over{x-1}}\right)
\right]\nonumber\\
&&\Pi_{_{L}}(l,l_0)=3 m^2_{\hbox{\cmr g}}\left[
(1-x^2)-{{x(1-x^2)}\over{2}}\ln\left({{x+1}\over{x-1}}\right)
\right]
\label{gluonself}
\end{eqnarray}
where $x\equiv l_0/l$ and $m^2_{\hbox{\cmr g}}\equiv g^2 T^2 [N+
N_{\hbox{\cmr f}}/2]/9$ is the gluon thermal mass in a $SU(N)$ gauge 
theory with $N_{\hbox{\cmr f}}$ flavors. Since these functions
depend only of $x=l_0/l$, we will often simplify the notation
 by simply indicating $x$ in their list of arguments.

It will be useful to
introduce the spectral functions
\begin{equation}
\rho_{_{T,L}}(l,l_0)\equiv\left.{{i}\over
{L^2-\Pi_{_{T,L}}(L)}}\right|_{_{R}}
-\left.{{i}\over{L^2-\Pi_{_{T,L}}(L)}}\right|_{_{A}} .
\end{equation}
The properties of these spectral functions are closely related to the
analytic structure of the gluon propagator. In particular, this 
propagator possesses two thermal mass-shells (a transverse one and a 
longitudinal one) above the light-cone, determined by the
equation $L^2=\Pi_{_{T,L}}(L)$, and to which correspond 
delta functions in 
$\rho_{_{T,L}}$
 for $L^2 > 0$. When $L^2< 0$ ($i.e.$ $|x|<1$),
 the self-energies
$\Pi_{_{T,L}}(L)$ acquire an imaginary part due to the
logarithm (a phenomenon known as Landau damping), so 
that the corresponding
contribution in $\rho_{_{T,L}}$ is:
\begin{equation}
{{-2\hbox{\rm Im}\,\Pi_{_{T,L}}(L) |_{_{R}}}\over
{\left(L^2-\hbox{\rm Re}\,\Pi_{_{T,L}}(L)\right)^2
+\left(\hbox{\rm Im}\,\Pi_{_{T,L}}(L)|_{_{R}}\right)^2}}.
\label{landaudamping}
\end{equation}
For completeness we give here the value of the imaginary parts
to be used later:
\bea
{\hbox{\rm Im}\,\Pi_{_{T}}(L)|_{_{R}}} &=& {3\pi m^2_{\hbox{\cmr g}} 
\over 4}
\ x (1-x^2)  \nonumber \\
{\hbox{\rm Im}\,\Pi_{_{L}}(L)|_{_{R}}} &=& - 2 \ 
{\hbox{\rm Im}\,\Pi_{_{T}}(L)|_{_{R}}} .
\label{impart}
\ena

\subsubsection{Resummed hard fermion propagator}
Since the calculation done with massless fermions leads to 
collinear divergences \cite{letter}, it has been proposed in 
\cite{rebhan}
to cure these singularities by taking into account the thermal
mass even for hard particles. Without any approximation, this
resummed propagator reads \cite{klim,pisars,wel2}:   
\begin{equation}                      
{\cal S}(P)={i\over 2}\left[
{{\gamma^0-\hat{p}\cdot\vec{\gamma}}\over{(p_0-\Sigma_0(P))
-\parallel\vec{p}-\vec{\Sigma}(P)\parallel+i\epsilon}}
+{{\gamma^0+\hat{p}\cdot\vec{\gamma}}\over{(p_0-\Sigma_0(P))
+\parallel\vec{p}-\vec{\Sigma}(P)\parallel-i\epsilon}}
\right],
\end{equation}
where, at the one loop order,
\begin{eqnarray}
&&\Sigma_0(P)={{M^2_\infty}\over{4p}}\ln\left({{x+1}
\over{x-1}}\right)\nonumber\\
&&\vec{\Sigma}(P)=\hat{p}{{M^2_\infty}\over{2p}}\left[
1-{x\over 2}\ln\left({{x+1}\over{x-1}}\right)\right]
\end{eqnarray}
with $x\equiv p_0/p$, and where $M^2_\infty\equiv
g^2 C_{_{F}} T^2/4$ stands
for the asymptotic fermion thermal mass.

It is well known that this resummed propagator possesses 
two mass-shells
(usually denoted by $+$ and $-$), the energy of which can 
be approximated for hard $p$ by:
\begin{eqnarray}
&&\omega_{+}(p)\build{p\sim T}\over{\approx}
\sqrt{p^2+M^2_\infty}\nonumber\\
&&\omega_{-}(p)\build{p\sim T}\over{\approx}
p\left(1+2\exp\left(-4p^2/M^2_\infty-1\right)\right)
\end{eqnarray}
Moreover, the residue associated to the $-$ pole is exponentially small
whereas the residue associated to the $+$ pole is approximately $1$.
This means that in a momentum integral involving such a
propagator and where $p$ is hard, we need only to take into account
the $+$ pole. In such circumstances, the resummed fermion propagator
can be simplified to get:
\begin{equation}
{\cal S}(P)\approx i\;{{p_0\gamma^0-\omega_{+}(p)\hat{p}\cdot
\vec{\gamma}}
\over{p_0^2-\omega^2_{+}(p)+i\epsilon}}
\approx {\slP}\;S(P),
\end{equation}
where
\begin{equation}
S(P)\equiv{i\over{P^2-M^2_\infty+i\epsilon}} .
\end{equation}

\subsection{Expression of $\mbox{\rm Im}\Pi^\mu\,_\mu$}
Armed with these tools, one can now evaluate the imaginary part of the
diagrams represented on Fig.~\ref{fighard}. The imaginary part of the photon
self-energy is related to the retarded and advanced ones by
\begin{equation}
2i\,\hbox{\rm Im}\,\Pi^\mu{}_\mu=\Pi^\mu{}_\mu{}_{|_{R}}-
\Pi^\mu{}_\mu{}_{|_{A}}
\end{equation}
In order to simplify the expressions, 
we will forget for a while all the color and group factors, and
reintroduce them only at the end of the calculation. 


Let us begin by the self-energy correction of Fig.~\ref{fighard}-(b) 
(keeping in mind
that exists a second such diagram). The contribution of this
diagram to the retarded self-energy of the photon is \cite{pat,ra}:
\begin{eqnarray} 
 -{i\Pi^\mu\,_\mu(Q)}_{|_{R}}^{\hbox{\cmr self}}  &=&  e^2 
\int {d^nR\over(2\pi)^n} \nonumber\\
  & &\times\left\{\left[\frac{1}{2}-n_{_{F}}(p_0)\right]
  \left[S_{_{R}}(P)-S_{_{A}}(P)\right]
\left[S_{_{R}}(R)\right]^2
  \hbox{\rm Tr}{\imb\Sigma}_{_{R}} \right.\nonumber\\
  & &+\left. \left[\frac{1}{2}-n_{_{F}}(r_0)\right]\left[
  S_{_{R}}^2(R)\hbox{\rm Tr}{\imb\Sigma}_{_{R}}-
S_{_{A}}^2(R)\hbox{\rm Tr}{\imb\Sigma}_{_{A}}\right]
  S_{_{A}}(P)\right\}
\end{eqnarray}
where the notation $\hbox{\rm Tr}{\imb\Sigma}_\alpha$
with $\alpha=R,\,A$ stands for:
\begin{equation}
  \hbox{\rm Tr}{\imb\Sigma}_\alpha\equiv\hbox{\rm Tr}\left(\gamma_\mu 
\slR
  \left[-i\Sigma_\alpha(R)\right] \slR \gamma^\mu \slP\right).
\end{equation}
A similar relation is easily  derived for the advanced self-energy.
The one-loop fermion self-energy $\Sigma_\alpha(R)$ is calculated in 
a covariant gauge with the effective soft gluon propagator of
Eq.~(\ref{gluon}). The thermal distributions can be arranged in a
simple way in order to give:
\begin{eqnarray}
&\hbox{\rm Im}\,\Pi^\mu{}_\mu(Q)^{\hbox{\cmr self}}&={1\over 2}
e^2 g^2 \int {{d^nR}\over{(2\pi)^n}}
\left[n_{_{F}}(r_0)-n_{_{F}}(p_0)\right]\nonumber\\
&&\times\int{{d^nL}\over{(2\pi)^n}}
\left[\Delta_{_{R}}^{^{T,L,G}}(L)-\Delta_{_{A}}^{^{T,L,G}}(L)\right]
P^{\rho\sigma}_{_{T,L,G}}(L)\;
\hbox{\rm Tr}^{\hbox{\cmr self}}_{\rho\sigma}\nonumber\\
&&\times
 \left[S_{_{R}}(P)-S_{_{A}}(P)\right]
\left[S_{_{R}}(R+L)-S_{_{A}}(R+L)\right]\nonumber\\
&&
\times\left[n_{_{B}}(l_0)+n_{_{F}}(r_0+l_0)\right]
{\cal P}\left({1\over{(R^2-M^2_\infty)^2}}\right)
\label{self}
\end{eqnarray}
for the transverse ($T$), longitudinal ($L$), and gauge ($G$) contributions
corresponding to the propagator of Eq.~(\ref{gluon}), 
where ${\cal P}$ denotes the principal value and 
$\hbox{\rm Tr}^{\hbox{\cmr self}}_{\rho\sigma}$ stands 
for the Dirac's trace
\begin{equation}
\hbox{\rm Tr}^{\hbox{\cmr self}}_{\rho\sigma}\equiv\hbox{\rm Tr}\,
\left[\gamma^\mu \slR \gamma_\sigma (\slR+\slL)\gamma_\rho\slR
\gamma_\mu\slP\right].
\end{equation}
The reason why we kept only the cut $(b)$ of Fig.~\ref{fighard}-(b) is
that when $Q^2\geq 0$ (which is the case of interest if one wants
to study the production rate of real photons or of dileptons), 
the cuts $(a)$ and $(c)$ are not allowed by kinematical constraints.
Indeed, an on-shell fermion can emit by bremsstrahlung only 
photons with a virtuality $Q^2<0$.


The same kind of calculations can be carried out for the contribution
of the vertex correction of Fig.~\ref{fighard}-(a) to the retarded
self-energy of the photon \cite{ra,pat}:
\begin{eqnarray}
&&\!\!\!\!\!\!\!\!-i{\Pi^\mu\,_\mu(Q)}_{|_{R}}^{\hbox{\cmr vertex}} 
 =  e^2 
\int {d^nR\over(2\pi)^n} \ 
P^{\rho\sigma}_{_{T,L,G}}(L)\;
\hbox{\rm Tr}_{\rho\sigma}^{\hbox{\cmr vertex}} \nonumber\\
 &&\!\!\!\!\!\!\!\!\!\!\!\! \times\left\{\left[\frac{1}{2}-
n_{_{F}}(p_0)\right]
  \left( V^{^{T,L,G}}_{_{RRA}}(P,Q,-R)S_{_{R}}(P) - 
          V^{^{T,L,G}}_{_{ARA}}(P,Q,-R)S_{_{A}}(P) \right) S_{_{R}}(R) 
          \right.\nonumber\\       
 &&\!\!\!\!\!\!\!\!\!\!\!\!\!\! + \left.\left[\frac{1}{2}-
n_{_{F}}(r_0)\right]
 \left( V^{^{T,L,G}}_{_{ARA}}(P,Q,-R)S_{_{R}}(R) - 
          V^{^{T,L,G}}_{_{ARR}}(P,Q,-R)S_{_{A}}(R) \right) S_{_{A}}(P) 
\right\}
\label{imvertex} 
\end{eqnarray}
again corresponding to the transverse ($T$), longitudinal ($L$), 
and gauge ($G$) terms of Eq.~(\ref{gluon}),
where $\hbox{\rm Tr}_{\rho\sigma}^{\hbox{\cmr vertex}}$ contains the 
Dirac's algebra factors:
\begin{equation}
\hbox{\rm Tr}_{\rho\sigma}^{\hbox{\cmr vertex}}\equiv
\hbox{\rm Tr}\left[\gamma_\mu\slR\gamma_\rho(\slR+\slL)
\gamma^\mu(\slP+\slL)
\gamma_\sigma\slP\right]
\end{equation}
and the functions $V^{^{T,L,G}}_{\alpha\beta\gamma}(P,Q,-R)$ are the
scalar part of the one loop $q-q-\gamma$ vertex, where the gluon line
of the loop is made of an effective propagator. These are equal to
\begin{eqnarray}
&&\!\!\!\!\!\!\!\!V^{^{T,L,G}}_{\alpha\beta\gamma}(P,Q,-R)=
- g^2  
\int {d^nL\over(2\pi)^n} \nonumber \\
&&\!\!\!\!\!\!\!\!\times\left\{ ( \frac{1}{2}+n_{_{B}}(l_0) )
    \left[\Delta^{^{T,L,G}}_{_{R}}(L)-
\Delta^{^{T,L,G}}_{_{A}}(L) \right] 
     S_\alpha(P+L) S_{\bar\gamma}(R+L) \right. \nonumber \\
&&\!\!\!\!\!\!\!\!+ ( \frac{1}{2}-n_{_{F}}(r_0+l_0) )
   \left[ S_{_R}(R+L) - S_{_A}(R+L) \right]
   S_{\bar\beta}(P+L) \Delta^{^{T,L,G}}_{\gamma}(L)
   \nonumber  \\
&&\!\!\!\!\!\!\!\!\left. + ( \frac{1}{2}-n_{_{F}}(p_0+l_0) )
  \left[ S_{_R}(P+L) - S_{_A}(P+L) \right]
   S_{\beta}(R+L)  \Delta^{^{T,L,G}}_{\bar\alpha}(L)
   \right\}. 
\label{vert}    
\end{eqnarray}
Inserting this expression into Eq.~(\ref{imvertex}), and calculating in 
the same way the advanced self-energy, one obtains the contribution of
Fig.~\ref{fighard}-(a) to the imaginary part of the photon 
self-energy:
\begin{eqnarray}
&\hbox{\rm Im}\,\Pi^\mu{}_\mu(Q)^{\hbox{\cmr vertex}}&={1\over 2}
e^2 g^2 \int {{d^nR}\over{(2\pi)^n}}
\left[n_{_{F}}(r_0)-n_{_{F}}(p_0)\right]\nonumber\\
&&\times\int{{d^nL}\over{(2\pi)^n}}
\left[\Delta_{_{R}}^{^{T,L,G}}(L)-\Delta_{_{A}}^{^{T,L,G}}(L)\right]
P^{\rho\sigma}_{_{T,L,G}}(L)\;
\hbox{\rm Tr}^{\hbox{\cmr vertex}}_{\rho\sigma}\nonumber\\
&&\times\left\{
\vphantom{{\cal P}\left({1\over{P^2-M^2_\infty}}\right)} 
\left[S_{_{R}}(P)-S_{_{A}}(P)\right]
\left[S_{_{R}}(R+L)-S_{_{A}}(R+L)\right]\right.\nonumber\\
&&\times\left[n_{_{B}}(l_0)+n_{_{F}}(r_0+l_0)\right]
{\cal P}\left({1\over{R^2-M^2_\infty}}\right)\;
{\cal P}\left({1\over{(P+L)^2-M^2_\infty}}\right)\;\nonumber\\
&&+\left[S_{_{R}}(R)-S_{_{A}}(R)\right]
\left[S_{_{R}}(P+L)-S_{_{A}}(P+L)\right]\nonumber\\
&&\left. \times\left[n_{_{B}}(l_0)+n_{_{F}}(p_0+l_0)\right]
{\cal P}\left({1\over{P^2-M^2_\infty}}\right)\;
{\cal P}\left({1\over{(R+L)^2-M^2_\infty}}\right)\;
\right\}
\label{vertex}
\end{eqnarray}
For the same kinematical reasons as in the self-energy diagram, 
the cuts $(a)$ and $(c)$ do not contribute to the production rate 
of real photons or lepton pairs.
We notice at this point a high symmetry between the
self-energy correction and the vertex correction.
Two kind of terms appear in the latter
 and can be interpreted in terms of cut
diagrams: the first term in the curly brackets above corresponds to
cut $(b)$ and is to be combined with Eq.~(\ref{self}) while the second
term should be combined with the other self-energy correction mentioned
above but not made explicit. Both classes of terms give identical
contributions to the photon production rate, so that in the following
we consider only the first term 
of Eq.~(\ref{vertex}) and Eq.~(\ref{self}), and take the
other terms into account simply by multiplying by
an overall factor $2$.

\subsection{Reduction of the traces}

In order to make some simplifications more obvious, it is useful
to compute the Dirac's traces 
$\hbox{\rm Tr}^{\hbox{\cmr self}}_{\rho\sigma}$
and
$\hbox{\rm Tr}_{\rho\sigma}^{\hbox{\cmr vertex}}$
in such a way that the invariants $P^2$, $R^2$, $(P+L)^2$ and 
$(R+L)^2$ appear whenever possible. We get:
\begin{eqnarray}
\hbox{\rm Tr}^{\hbox{\cmr self}}_{\rho\sigma}=-4\left[
4R^2 Q_\rho R_\sigma -4Q^2 R_\rho R_\sigma 
-g_{\rho\sigma}\left(
R^2(R^2-Q^2)+2R^2 Q\cdot L-2Q^2 R\cdot L 
\right)\right]
\end{eqnarray}
and 
\begin{eqnarray}
&\hbox{\rm Tr}_{\rho\sigma}^{\hbox{\cmr vertex}}=-4&\left[
2(R^2+(R+L)^2) P_\rho Q_\sigma -2(P^2+(P+L)^2) R_\rho Q_\sigma
\right.\nonumber\\
&&+2L^2(R_\rho R_\sigma+P_\rho P_\sigma)-4Q^2 R_\rho P_\sigma
\nonumber\\
&\qquad\qquad +g_{\rho\sigma}&\left(
P^2R^2+(P+L)^2(R+L)^2\right.\nonumber\\
&&\left.\left.-L^2(P^2+R^2+(P+L)^2+(R+L)^2-Q^2-L^2)
\right)\right].
\end{eqnarray}
At this point, it is worth noting that we dropped terms 
proportional 
to $L_\rho$ or $L_\sigma$, since they will disappear in the 
contraction with the projectors $P_{_{T,L}}^{\rho\sigma}$:
\begin{equation}
L_\rho\;P^{\rho\sigma}_{_{T,L}}(L)=0.
\end{equation}
Therefore, the expressions obtained above for the Dirac's 
traces should not be
used to compute the gauge dependence of the production rate, 
since
$L_\rho\;P^{\rho\sigma}_{_{G}}(L)\not=0$.  In order to verify 
the independence
of this rate with respect to the gauge parameter $\xi$, it is 
simpler to go
back to the original expressions Eqs.~(\ref{self}),~(\ref{vertex}) 
and show a
compensation of the gauge dependence between the self-energy and 
the vertex
contributions, exactly in the same manner as at $T=0$.

\subsection{Kinematics}
We now take into account the constraints provided by the
discontinuities:
\begin{eqnarray}
&&S_{_{R}}(P)-S_{_{A}}(P)=2\pi\epsilon(p_0)\delta(P^2-M^2_\infty)
\nonumber\\
&&S_{_{R}}(R+L)-S_{_{A}}(R+L)=2\pi\epsilon(r_0+l_0)
\delta((R+L)^2-M^2_\infty);
\end{eqnarray}
this will enable us to perform two of the momentum
integrals ``for free". Moreover, as already mentioned, the components
of $R$ and $P$ are considered to be hard, whereas the
components of $Q$ and $L$ are considered to be much smaller than
the temperature $T$.
Therefore, $p_0$ and $r_0+l_0$ are of the same sign:
\begin{equation}
\epsilon(p_0)\epsilon(r_0+l_0)=1 .
\end{equation}

From the constraint $\delta(P^2-M^2_\infty)$, we extract the
value of $r_0$:
\begin{eqnarray}
&r_0&=q_0\pm \sqrt{r^2-2 q r \cos\theta +q^2+M^2_\infty}\nonumber\\
&&\approx q_0\pm\left[r-q\cos\theta+{{q^2}\over{2r}}(1-\cos^2\theta)
+{{M^2_\infty}\over{2r}}\right],
\label{r0equal}
\end{eqnarray}
where $\theta$ is the angle between the spatial components of $R$ and
$Q$.
 Since the possible two signs for $r_0$ lead at
the end to equivalent contributions, we will 
consider only the positive 
solution for $r_0$, and multiply when necessary the result by a factor 
$2$ to take into account the other solution.
From the constraint $\delta((R+L)^2-M^2_\infty)$, we obtain the
value of $\cos\theta^\prime$, where $\theta^\prime$ is the angle between
the spatial components of $R$ and $L$:
\begin{eqnarray}
&\cos\theta^\prime&={{(r_0+l_0)^2-(r^2+l^2+M^2_\infty)}\over{2rl}}
\nonumber\\
&&\approx {{l_0}\over{l}}+{{q_0-q\cos\theta}\over{l}}.
\label{constraint}
\end{eqnarray}
Moreover, we must impose $\cos\theta^\prime$ to be in $[-1,1]$,
which can give additional constraints between $l_0,l,q_0,q$ and 
$\cos\theta$.
When needed, the variable $\cos\theta^{\prime\prime}$, where
$\theta^{\prime\prime}$ is the angle between $Q$ and $L$, will be
obtained by:
\begin{equation}
\cos\theta^{\prime\prime}=\cos\theta\cos\theta^\prime+\sin\theta
\sin\theta^\prime\cos\phi,
\label{angle}
\end{equation}
where $\phi$ is the angle between the projections of $Q$ and $L$
on the plane defining the spherical coordinates basis.
Given these relations,
 the angular variables we choose as independent ones are
$\theta$ and $\phi$, $\theta^\prime$ being extracted from the
constraint Eq.~(\ref{constraint}) and $\theta^{\prime\prime}$ 
being given by Eq.~(\ref{angle}).

With such a choice of independent variables, we can calculate
approximate values for the various denominators entering the rate:
\begin{equation}
R^2-M^2_\infty=2Q\cdot R -Q^2\approx 2q_0 r\left[1-\cos\theta
+{{M^2_\infty}\over{2r^2}}+{{Q^2}\over{2q_0^2}}\right],
\label{deno1}
\end{equation}
\begin{eqnarray}
&\displaystyle{\int\limits_{0}^{2\pi}
{{d\phi}\over{(P+L)^2-M^2_\infty}}}
&=\int\limits_{0}^{2\pi}{{d\phi}\over{Q^2-2Q\cdot R -2 Q\cdot L}}
\nonumber\\
&&\approx {{2\pi}\over{2q_0 r\left[
\left(1-\cos\theta+
{{M^2_\infty}\over{2r^2}}+{{Q^2}\over{2q_0^2}}+
{{L^2}\over{2r^2}}\right)^2-{{L^2}\over{r^2}}
\left({{M^2_\infty}\over{r^2}}+{{Q^2}\over{q_0^2}}\right)
\right]^{1/2}}},
\label{deno2}
\end{eqnarray}
(the reason why we performed the integration over $\phi$ at this point 
is that at the dominant order $(P+L)^2-M^2_\infty$ is the only
quantity in which $\phi$ appears).

It is worth noticing that with $M_\infty=0$ and $Q^2=0$, the
denominator $R^2-M^2_\infty$ exhibits a collinear singularity
when $\theta=0$, $i.e.$ when the photon is emitted collinearly
to the hard fermion. Looking at the previous expressions, we see
that it is natural to introduce the new variables:
\begin{eqnarray}
&&u\equiv 1-\cos\theta\nonumber\\
&&M^2_{\hbox{\cmr eff}}\equiv M^2_\infty+{{Q^2r^2}\over{q_0^2}}.
\label{masseff}
\end{eqnarray}
It is now clear that the collinear divergence is regulated by the
effective fermion mass $M_{\hbox{\cmr eff}}$ which acts as
a cut-off in the integral over $u$. Therefore, the order of the
result depends on the order of this cut-off, a question 
which will be discussed
in the following paragraph.

\subsection{Enhancement mechanism}
\label{enhancement}

{\it A priori}, 
we are faced with four kinds of angular integrals over the
variable $u$:
\begin{eqnarray}
&\displaystyle{I_1\equiv\int\limits_{0}^{2}
{{du}\over{R^2-M^2_\infty}},\;} \qquad
&I_2\equiv\int\limits_{0}^{2}{{du}\over{(P+L)^2-M^2_\infty}},\nonumber\\
&\displaystyle{I_3\equiv\int\limits_{0}^{2}
{{du}\over{(R^2-M^2_\infty)^2}},\;}\qquad
&I_4\equiv\int\limits_{0}^{2}{{du}\over{(R^2-M^2_\infty)(
(P+L)^2-M^2_\infty)}}.
\end{eqnarray}
Since the order of $R^2-M^2_\infty$ or $(P+L)^2-M^2_\infty$ is
$q_0 r$, the orders we expect na\"\i vely for these integrals
are:
\begin{eqnarray}
&&I_1, I_2={{\cal O}}\left({1}\over{q_0 r}\right) \nonumber\\
&&I_3, I_4={{\cal O}}\left({1}\over{(q_0 r)^2}\right); 
\end{eqnarray} 
and this would be right, up to logarithmic factors, if the collinear 
singularities where really logarithmic in the four integrals.
This is actually the case for $I_1$ and $I_2$ but obviously
not for $I_3$, which exhibits a power-like collinear singularity.
Looking at the form of $R^2-M^2_\infty$, we have for the order of
$I_3$:
\begin{equation}
I_3={{\cal O}}\left({{1}\over{(q_0 r)^2}}\;{{1}\over{u^*}}\right),
\end{equation}
where $u^*\sim M^2_{\hbox{\cmr eff}}/r^2$. Therefore, the actual order
of $I_3$ depends on the order of the physical cut-off in the $u$ 
integral, this dependence being a power-like one instead of
a logarithmic one. Moreover, this integral is enhanced with 
respect to its
na\"\i ve order when $u^*\ll1$, which occurs when the virtuality 
of the emitted photon is small.

As for $I_4$, since the
hard momenta $R$ and $P+L$ only differ by soft quantities, 
the denominators $R^2-M^2_\infty$ and $(P+L)^2-M^2_\infty$ 
are zero almost simultaneously.
Let us quantify more precisely this assertion; to explain 
roughly what
happens in this case, it is sufficient to write
$R^2-M^2_\infty\sim q_0 r(u+u^*)$ and $(P+L)^2-M^2_\infty
\sim q_0 r(u+u^{*\prime})$. Again, we
have $u^*\sim M^2_{\hbox{\cmr eff}}/r^2$. Moreover, 
 $u^{*\prime}-u^*$ is independent of the value of 
$M^2_{\hbox{\cmr eff}}/r^2$ and is of order $L^2/r^2\ll 1$. Therefore, 
we get the following orders for the integral $I_4$:
\begin{eqnarray}
&&\hbox{\rm If\ \ }u^*\le L^2/r^2: \qquad I_4={{\cal O}}\left(
{{1}\over{(q_0 r)^2}}
{{1}\over{u^{*\prime}-u^*}}\right)\gg {{1}\over{(q_0 r)^2}}\nonumber\\
&&\hbox{\rm If\ \ }u^*\gg L^2/r^2: \qquad I_4={{\cal O}}
\left({1\over{(q_0 r)^2}}
{{1}\over{u^*}}\right)
\end{eqnarray}
Like $I_3$, this integral is enhanced when $u^*\ll 1$. 
The reason why $I_4$
behaves much like $I_3$ whereas $I_4$ contains two simple poles
instead of a double one in the case of $I_3$ lies, of course, in 
the very
close vicinity of these two poles. In fact, when $u^*$ is smaller
than the distance between the two poles, the order of the result is 
controlled by this separation which is of order $L^2/r^2$. On the
contrary, when the regulator $u^*$ is larger than this separation, 
everything happens as if we had a double pole.

Recalling now that the fermionic thermal mass $M_\infty$ is of order 
$gT$, we should distinguish various cases according to the value of the 
virtuality $Q^2$ (see Eq.~(\ref{masseff})): \label{cases}
\begin{itemize}
\item[(i)] $Q^2/q_0^2 \ll g^2\, :$ in this case, the virtuality 
of the emitted photon does not play any role, and can be completely 
neglected, since $M_{\hbox{\cmr eff}}\approx M_\infty$. This case
will be studied first since it is the simplest one, and will provide
a basis for the case where the virtuality can no longer be neglected.
In this case, the integrals $I_3$ and $I_4$ are enhanced whereas
$I_1$ and $I_2$ have their ``normal" order, therefore we can 
neglect every occurrence of $I_1$ or $I_2$.

\item[(ii)] $g^2 \le Q^2/q_0^2 \ll 1\, :$ in this case, the 
virtuality must be taken into account since its effect
is at least as important as the effect of the fermion thermal mass.
We still have $u^*\ll1$, so that the remark made in (i) on
the enhancement of $I_3$ and $I_4$ remains true. This section
will make an intensive use of the results obtained 
in the case (i).

\item[(iii)] $Q^2/q_0^2\sim 1\, :$ here also, the virtuality of the
photon plays an important role. Moreover, this case is very different
from the previous two since none of the $I_i$'s is enhanced, and therefore
all must be taken into account. Another reason why this
case is very different lies in the fact that there is no more a
hierarchy between the various powers of $Q^2/q_0^2$; as a consequence,
all the powers of this quantity should be kept in the expansion of the 
numerator. 
\end{itemize}

It is interesting to compare the status of mass singularities in our calculation
with results obtained in previous works.  Collinear singularities
and their cancelation have been studied in (non-resummed) perturbation theory
at two-loop order in QED/QCD \cite{collin,collin1} and three-loop order in the
$\lambda \phi^3$ model \cite{collin2}:  in all cases the problem considered was
the decay rate of a heavy particle into massless particles and a complete
cancelation of collinear singularities associated to the massless particle
propagators was found when summing over real and virtual diagrams and over all
thermal processes.  The equivalent problem here is the case $Q^2 \ne 0$ but
$M^2_\infty = 0$ and indeed our expressions are regulated by the photon
virtuality:  in the R/A formalism there is no need to distinguish between
``real" and ``virtual" diagrams as they are all included in rather compact
expressions (see $e.g.$ Eq.~(\ref{sum}) below).  It is the masslessness of the
``decaying" particle ($Q^2 = 0$) which generates the collinear singularity of
interest here as it would in the previous studies if we had let $Q^2
\rightarrow 0$:  indeed the two-loop correction to 
the invariant rate $R$  was found to be $\sim e^2 g^2 T^2 
\ln (T^2 / Q^2)$
\cite{collin3}, which is indeed logarithmically divergent when the photon virtuality vanishes.
This softer singularity (compared to the ``power-like" divergence above) 
can be understood because no HTL resummation was performed on
the gluon propagator ($L^2 = 0$) and therefore no Landau damping was included.

\vskip 1cm\goodbreak
\section{Production of quasi real photons: $Q^2/{\lowercase{q}_0^2}\ll 
{\lowercase{g}^2}$}
Let us now specialize to the production rate of almost real photons,
which corresponds to the case denoted by (i) at the end of the
previous section. From a 
technical point of view, this case is the simplest one. Nevertheless,
 it
will serve as a basis for the production rate of photons having a small
virtuality, since this latter case will appear as a 
generalization
of the present case.
\subsection{Expression of $\mbox{\rm Im}\Pi^\mu\,_\mu$}
By using the fact that $Q^2=0$ and the fact that
 $P,R$ are hard and $Q,L$ are 
much smaller than $T$, it is possible to greatly simplify
the expression of the imaginary part of the photon self-energy. In
particular, we get some partial cancelations between the contributions 
of the self-energy and of the vertex. The remaining dominant term is:
\begin{eqnarray}
 & & \hbox{\rm Im}\,\Pi^\mu\,_\mu(Q)\approx -8e^2g^2
\int\frac{d^n R}{(2\pi)^{n-1}}
  \int\frac{d^n L}{(2\pi)^{n-1}}
  {q_0} n_{_{F}}'(p_0)\; n_{_{B}}(l_0)\rho_{_{T,L}}(l,l_0) \nonumber\\
  & &\qquad\times 
  \epsilon(p_0)\epsilon(r_0+l_0)\delta(P^2-M^2_\infty)
\delta\left[(R+L)^2-M^2_\infty\right]\nonumber\\
& &\qquad\times 
  (R_\rho R_\sigma+P_\rho P_\sigma) P_{_{T,L}}^{\rho\sigma}
  \frac{L^2}{(R^2-M^2_\infty)((P+L)^2-M^2_\infty)},
\label{sum}
\end{eqnarray}
where we took into account the fact that the integral $I_4$ is 
enhanced. We have also simplified the statistical 
factor $n_{_{F}}(r_0) - n_{_{F}}(p_0)$ to $q_0 n_{_{F}}'(p_0)$ 
which is justified
within our kinematical approximation. The fact that the first 
non vanishing term in the numerator is proportional to $L^2 \ll T^2$
shows clearly that ${\rm Im}\,\Pi^\mu{}_\mu(Q)$ is zero at the HTL
order.

Using Eqs.~(\ref{r0equal}), (\ref{constraint}), the expansion of the 
terms appearing in the numerator gives
\begin{eqnarray}
&L^2 (R_\rho R_\sigma+P_\rho P_\sigma) P_{_{T}}^{\rho\sigma}&\approx
2 \left({{r}\over{l}}\right)^2 (L^2)^2\nonumber\\
&L^2 (R_\rho R_\sigma+P_\rho P_\sigma) P_{_{L}}^{\rho\sigma}&\approx
-2 \left({{r}\over{l}}\right)^2 (L^2)^2.
\label{numerator}
\end{eqnarray}
Using now the constraint $-1\leq \cos\theta'\leq 1$ 
(Eq.~(\ref{constraint})), we see that the gluon momentum $L$ is constrained to 
satisfy $l_0/l\in[-1,1]$, $i.e.$ $L^2<0$. 
This means that the production rate of real photons is 
dominated by the exchange of a space-like gluon (i.e., by
the Landau damping part of the gluon spectral density).
Eq.~(\ref{sum}) can be then be put into the form
\begin{eqnarray}
&\hbox{\rm Im}\,\Pi^\mu\,_\mu(Q)&\approx
(-1)_{_{T}}{{e^2 g^2}\over{8\pi^4}}
{1\over{q_0}}\int\limits_{r^*}^{\infty}dr n_{_{F}}^\prime(r)
\ \int\limits_{0}^{l^*}l^4 dl \int_{-1}^{+1}{dx}\,n_{_{B}}(lx)\, 
\rho_{_{T,L}}(l,lx)(1-x^2)^2\nonumber\\
&& \qquad \qquad \times\int\limits_{0}^{2} {{du}\over{\left[u+{{M_\infty^2}\over{2r^2}}
\right]\left[\left(u+{{M_\infty^2}\over{2r^2}}+{{l^2(x^2-1)}\over{2r^2}}
\right)^2+{{l^2(1-x^2)M_\infty^2}\over{r^4}}\right]^{1/2}}},
\label{imreal}
\end{eqnarray}
where we denoted $x\equiv l_0/l$, and where the symbol $(-1)_{_{T}}$
denotes an extra minus sign in the transverse contribution. 
 We have introduced some cut-offs $r^*$ and $l^*$ 
at a scale intermediate between $gT$ and $T$, since we
{\it a priori} restricted $r$ to be hard and $l$ to be negligeable in front of $T$. 
We will see later that these cut-offs can be taken respectively to
$0$ and $+\infty$, without modifying significantly the result. 

An alternate expression for $\hbox{\rm Im}\,\Pi^\mu\,_\mu(Q)$,
which will be needed later, can be obtained in the following 
way. Using the $\delta-$function constraints one easily derives
\be
{1\over R^2-M^2_\infty} {1\over (P+L)^2-M^2_\infty} =
-{1\over 2 Q\cdot L} \left( {1\over R^2-M^2_\infty} + 
{1\over (P+L)^2-M^2_\infty} \right).
\ee
Then, noticing that the change of variables
$P+L\rightarrow -R$ leaves the integrand in Eq.~(\ref{sum}) invariant 
it is legitimate to replace in this equation 
$[(R^2-M^2_\infty )((P+L)^2-M^2_\infty)]^{-1}$ by 
$[-Q\cdot L(R^2-M^2_\infty)]^{-1}$. We are then lead to the 
expression \footnote{An alternative 
 method to obtain the same result
directly from Eq.~(\ref{imreal}) is 
to perform a few changes of variables:
\begin{eqnarray}
&&u\equiv {{lM_\infty}\over{r^2}}\sqrt{1-x^2}t+{{l^2(1-x^2)}\over
{2r^2}}
-{{M^2_\infty}\over{2r^2}} \nonumber\\
&&t\equiv{1\over2}\left({{\alpha_0}\over{\alpha}}-
{{\alpha}\over{\alpha_0}}\right)\quad\hbox{\rm with\ \ }
\alpha_0\equiv{{M_\infty}\over{l\sqrt{1-x^2}}}\nonumber\\
&&\alpha\equiv{{1-\sqrt{1-u^\prime}}\over{2}}.
\end{eqnarray}
}
\begin{eqnarray}
&\hbox{\rm Im}\,\Pi^\mu\,_\mu(Q)&\approx
(-1)_{_{T}}{{e^2 g^2}\over{2\pi^4}}
{1\over{q_0}}\int\limits_{r^*}^{\infty}r^2\,dr n_{_{F}}^\prime(r)
\nonumber\\
&&\times\int\limits_{0}^{l^*}l^4 dl \int_{-1}^{+1}{dx}\,
n_{_{B}}(lx)\, 
\rho_{_{T,L}}(l,lx)(1-x^2)^2\int\limits_{0}^{1} 
\frac{du^\prime}{\sqrt{1-u^\prime}}
\frac{1}{4M_\infty^2+l^2(1-x^2)u^\prime},
\label{imreal2}
\end{eqnarray}
where the relation between the $u'$ and $u$ integration variables
is $u' \equiv - 8 r^2 u /L^2$.
 
\subsection{Reduction of the expression}
\label{reduction}
The expression of Eq.~(\ref{imreal2}) can be simplified by performing
the angular integration over the variable $u^\prime$. 
This integral is elementary and gives:
\begin{equation}
\int\limits_{0}^{1}\frac{du^\prime}{\sqrt{1-u^\prime}}
\frac{1}{4M_\infty^2+l^2(1-x^2)u^\prime}
=2{{\hbox{\rm tanh}^{-1}\sqrt{-L^2/(4M^2_\infty-L^2)}}
\over{\sqrt{-L^2(4M^2_\infty-L^2)}}}.
\label{uinteg}
\end{equation}
At this point, we notice that the integral over $r$ can now be 
factorized in order to obtain:
\begin{equation}
\int\limits_{r^*}^{+\infty}dr r^2 n_{_{F}}^\prime(r)\approx
\int\limits_{0}^{+\infty}dr r^2 n_{_{F}}^\prime(r)=-{{\pi^2 T^2}
\over{6}}.
\end{equation}
The reason why we can set the cut-off $r^*$ to zero lies in the 
fact that the function to be integrated is peaked around $r\sim T$,
whereas $r^*\ll T$. Indeed:
\begin{equation}
\int\limits_{0}^{r^*}dr r^2 n^\prime_{_{F}}(r)\approx
 -{{(r^*)^3}\over{12T}}.
\end{equation}
Moreover, in order to further simplify the result, we will approximate
the Bose-Einstein weight by $n_{_{B}}(lx)\approx T/lx$ since
the momentum $l$ is assumed to be much less than $T$.

Finally, if we recall that the spectral density to be used here
is given by Eq.~(\ref{landaudamping}) since $L^2<0$, and if we 
introduce some convenient dimensionless quantities:
\begin{eqnarray}
&&w\equiv {{-L^2}\over{M^2_\infty}}\nonumber\\
&&\widetilde{I}_{_{T,L}}(x)\equiv {{\hbox{\rm Im}\,
\Pi_{_{T,L}}(x)}\over{M^2_\infty}}\nonumber\\
&&\widetilde{R}_{_{T,L}}(x)\equiv {{\hbox{\rm Re}\,
\Pi_{_{T,L}}(x)}\over{M^2_\infty}},
\end{eqnarray}
the imaginary part of the real photon self-energy can be written as:
\begin{eqnarray}
& & \hbox{\rm Im}\,\Pi^\mu\,_\mu(Q)\approx(-1)_{_{T}}
  \frac{e^2 g^2 N_{_{C}} C_{_{F}}}{3\pi^2}
  \frac{T^3}{q_0}\int\limits_{0}^{1}\;
\frac{dx}{x}\;\widetilde{I}_{_{T,L}}(x) \nonumber\\
& & \qquad\times\int\limits_{0}^{w^*}\;dw\;
\frac{\sqrt{w/(w+4)}
\hbox{\rm tanh}{}^{-1}\sqrt{w/(w+4)}}{(w+\widetilde{R}_{_{T,L}}(x))^2+
(\widetilde{I}_{_{T,L}}(x))^2},
\label{imag}
\end{eqnarray}
where we have re-introduced the required color and group factors.

\subsection{Reduction using sum rules}
\label{sumrules}
Instead of performing first the angular integral over $u$ to
reduce Eq.~(\ref{imreal}) to the two-dimensional result of
Eq.~(\ref{imag}), it is possible to use sum rules
to obtain the result as a one-dimensional integral. 

Let us first recall that sum rules come in this context from
the spectral representation of the resummed propagator,
 which can be
written for the scalar-like part of the gluon propagator as:
\begin{equation}
{{i}\over{l_0^2-l^2-\Pi_{_{T,L}}(l,l_0)+i\epsilon}}
=\int\limits_{-\infty}^{+\infty}dE\,{{E}\over{2\pi}}\;
\rho_{_{T,L}}(l,E)\;
{{i}\over{l_0^2-E^2+i\epsilon}}.
\end{equation}
By taking the imaginary part of this equation, keeping in 
mind that $\rho_{_{T,L}}$ are real functions, we can then obtain:
\begin{equation}
\int\limits_{-\infty}^{+\infty}{{dx}\over{2\pi}}\,x\;\rho_{_{T,L}}(l,lx)
\; {\cal P}\left({1\over{y^2-x^2}}\right)=
{{l^2(y^2-1)-\hbox{\rm Re}\Pi_{_{T,L}}(y)}
\over{\left(l^2(y^2-1)-\hbox{\rm Re}\Pi_{_{T,L}}(y)\right)^2+
\left(\hbox{\rm Im}\Pi_{_{T,L}}(y)\right)^2}},
\label{sumrulegeneral}
\end{equation}
where we denoted $l_0\equiv ly$ and $E\equiv lx$. By taking now 
special 
values of $y$, one easily obtains:
\begin{eqnarray}
&\hbox{\rm With\ }y=0:\ \ &
\int\limits_{-\infty}^{+\infty}{{dx}\over{2\pi}}
\;{{\rho_{_{T}}(l,lx)}\over{x}}={{1}\over{l^2}}\nonumber\\
&&\int\limits_{-\infty}^{+\infty}{{dx}\over{2\pi}}
\;{{\rho_{_{L}}(l,lx)}\over{x}}={{1}\over
{l^2+3m^2_{\hbox{\cmr g}}}}\nonumber\\
&\hbox{\rm With\ }y=\infty:\ \ &
\int\limits_{-\infty}^{+\infty}{{dx}\over{2\pi}}
\,x\;\rho_{_{T,L}}(l,lx)={{1}\over{l^2}}\nonumber\\
&\hbox{\rm With\ }|y|>1:\ \ &
\int\limits_{-\infty}^{+\infty}{{dx}\over{2\pi}}
\,x\;\rho_{_{T,L}}(l,lx)\;{\cal P}\left({1\over{y^2-x^2}}\right)
={{1}\over{l^2(y^2-1)-\hbox{\rm Re}\Pi_{_{T,L}}(y)}}.
\end{eqnarray}
These are the basic sum rules we will use to perform the integral over
$x$. However, the integrals we need to perform are on 
$x\in[-1,+1]$ whereas the above sum rules give the result for an
integration over the 
whole real axis. Therefore, it will be necessary to subtract 
the contributions of $x\in ]-\infty,-1]\cup[1,+\infty[$. This 
subtraction is easy to perform since in that region, the spectral 
functions $\rho_{_{T,L}}(l,lx)$ are made of delta functions 
corresponding to the thermal mass-shells of the gluon.
 More precisely, we have:
\begin{equation}
\int\limits_{]-\infty,-1]\cup[1,+\infty[}{{dx}\over{2\pi}}
\;\rho_{_{T,L}}(l,lx)\;
f(x)={{Z_{_{T,L}}}\over{2l\omega_{_{T,L}}}}
\left[f\left({{\omega_{_{T,L}}}
\over{l}}\right)-f\left(-{{\omega_{_{T,L}}}
\over{l}}\right)\right],
\label{subtraction}
\end{equation}
where $Z_{_{T,L}}$ stands for the residue of the spectral function
on the mass-shell, whose energy is $\omega_{_{T,L}}(l)$. Explicit
forms of these residues can be found in Ref.\cite{brat}; for later
use we record here the asymptotic limits
\begin{eqnarray}
&\!\!\!\! \displaystyle{\omega^2_{_{T}}(l\to\infty)\simeq l^2+\frac{3}{2}m^2_{\hbox{\cmr g}}},
\qquad & 
\omega^2_{_{L}}(l\to\infty)\simeq l^2\left[
1+4\exp\left(-\frac{2l^2}{3 m^2_{\hbox{\cmr g}}}-2\right)\right],\nonumber\\
&\!\!\!\! \displaystyle{Z_{_{T}}(l\to\infty)\simeq 1 - \frac{3 m^2_{\hbox{\cmr g}}}{4l^2}
\ln\left(\frac{4l^2}{3 m^2_{\hbox{\cmr g}}}\right)},\qquad &
Z_{_{L}}(l\to\infty)\simeq \frac{2l^2}{3 m^2_{\hbox{\cmr g}}}\left[1-{{4l^2}\over{m^2}}
\exp\left(-\frac{2l^2}{3 m^2_{\hbox{\cmr g}}}-2\right)\right],\nonumber\\
&\!\!\!\! \displaystyle{\omega^2_{_{T}}(l\to 0)\simeq m^2_{\hbox{\cmr g}}+\frac{6}{5}l^2},\qquad  & 
\omega^2_{_{L}}(l\to 0)\simeq m^2_{\hbox{\cmr g}}+\frac{3}{5}l^2,\nonumber\\
&\!\!\!\!\displaystyle{ Z_{_{T}}(l\to 0) \simeq 1-\frac{l^2}{5m^2_{\hbox{\cmr g}}}}, \qquad & 
Z_{_{L}}(l\to 0) \simeq 1+{{2l^2}\over{5m^2_{\hbox{\cmr g}}}}.
\label{asymptTL}
\end{eqnarray}
As well, we can introduce in a natural
way a magnetic mass for the transverse gluon by assuming that
it modifies the static behavior in such a way that:
\begin{equation}
\Pi_{_{T}}(0)=m^2_{\hbox{\cmr mag}},
\label{magmass1}
\end{equation}
which, thanks to Eq.~(\ref{sumrulegeneral}), is
equivalent to:
\begin{eqnarray}
  & &\int\limits_{-\infty}^{+\infty}{{dx}\over{2\pi}}
\,x\rho_{_{T}}(l, lx)=\frac{1}{l^2},\nonumber\\
  & &\int\limits_{-\infty}^{+\infty}\frac{dx}{2\pi}
{{\rho_{_{T}}(l, lx)}\over{x}}=
  \frac{1}{l^2+m_{\hbox{\cmr mag}}^2}.
\label{magmass}\end{eqnarray}

These sum rules may now be used to reduce Eq.~(\ref{imreal2})
down to a one--dimensional integral. To do this, we first
cast Eq.~(\ref{imreal2}) into the form
\begin{eqnarray}
&\hbox{\rm Im}\,\Pi^\mu\,_\mu(Q)&\approx
(-1)_{_{T}}{{e^2 g^2}\over{12\pi^2}}
{T^3\over{q_0}}
\,\int\limits_{0}^{l^*}l\,dl \int\limits_{-1}^{+1}dx\;
\rho_{_{T,L}}(l,lx)
\int\limits_{0}^{1} {{du^\prime}\over
{u^\prime\sqrt{1-u^\prime}}}\times\nonumber\\
&&\frac{1}{2y^2}\left\{
\frac{2}{x}(x^2-1)+2x(y^2-1)+(y^2-1)^2
\left[\frac{1}{x+y}+\frac{1}{x-y}\right]\right\}
\label{imsimple}
\end{eqnarray}
where $y\equiv \sqrt{1+4M^2_\infty/l^2u^\prime}$.
The application of sum rules to
evaluate Eq.~(\ref{imsimple}) is now completely straightforward.
We consider each of the terms in curly brackets in turn,
and examine first of all the transverse contribution.

For the first term, we have a contribution due to
\begin{eqnarray}
& &  K_1\equiv -2\int\limits_{0}^{l^*}l\,dl \int\limits_{-1}^{+1}dx\;
\rho_{_{T,L}}(l,lx)
\int\limits_{0}^{1} {{du^\prime}\over
{u^\prime\sqrt{1-u^\prime}}}\frac{1}{2y^2}
\frac{2}{x}(x^2-1)\nonumber\\
& &=\int\limits_{0}^{l^*}\frac{dl}{\sqrt{l^2+4M_\infty^2}}
\ln\left[\frac{(l+\sqrt{l^2+4M_\infty^2})^2}{4M_\infty^2}\right]
\left\{(1-Z_{_{T}})-l^2\left(\frac{1}{l^2+m_{\hbox{\cmr mag}}^2}-
\frac{Z_{_{T}}}{\omega_{_{T}}^2}\right)\right\},
\label{int1}\end{eqnarray}
where we have introduced a magnetic mass $m_{\hbox{\cmr mag}}\sim g^2T$
via Eq.~(\ref{magmass}).
Thanks to Eq.~(\ref{asymptTL}),
this integral is finite for $l\to 0$, and in fact for
$M_\infty\sim gT$ we can drop $m_{\hbox{\cmr mag}}\sim g^2T$,
but not if $M_\infty\sim g^2T$ or below. In the following
we use the HTL result, $M_\infty\sim gT$, so that we can drop
the dependence on the magnetic mass. It is also easy
to see from Eq.~(\ref{int1}) that since
\begin{equation}
  m^2_{\hbox{\cmr g}}\int\limits_{l^*}^{\infty}\frac{dl}{l^3}
  \ln\left(\frac{l}{M_\infty}\right) 
  \sim \left(\frac{m_{\hbox{\cmr g}}}{l^*}\right)^2 
  \ln\left(\frac{l^*}{M_\infty}\right),
\end{equation}
we can take $l^*\to\infty$ by introducing a negligeable 
contribution.

For the remaining terms of Eq.~(\ref{imsimple}), we find the
sum rules lead to
\begin{eqnarray}
 && K_2\equiv -2\int\limits_{0}^{l^*}l\,dl \int\limits_{-1}^{+1}dx\;
\rho_{_{T,L}}(l,lx)
\int\limits_{0}^{1} {{du^\prime}\over
{u^\prime\sqrt{1-u^\prime}}}\frac{1}{2y^2}
\left\{2x(y^2-1)+(y^2-1)^2
\left[\frac{1}{x+y}+\frac{1}{x-y}\right]\right\}.\nonumber\\
&&=\int\limits_{0}^{l^*}\frac{dl}{l}
\int\limits_{0}^{1} {{du^\prime}\over
{\sqrt{1-u^\prime}}}
\frac{4M_\infty^2}{4M_\infty^2+l^2u^\prime}
\left\{Z_{_{T}}\frac{\omega_{_{T}}^2-l^2}
  {4M_\infty^2-u^\prime(\omega_{_{T}}^2-l^2)}-
  \frac{{\rm Re}\,\Pi_{_{T}}(y)}
    {4M_\infty^2-u^\prime{\rm Re}\,\Pi_{_{T}}(y)}\right\}.
\label{int2}\end{eqnarray}
We note that no magnetic mass term arises for this contribution.
In this form, using
Eqs.~(\ref{asymptTL}),
 it is straightforward to verify that Eq.~(\ref{int2})
is finite as $l\to 0$, as there is a cancelation
between the two terms in curly brackets. As well, as was
the case of $K_1$, we can take the
limit of the cutoff $l^*\to\infty$ by introducing a
negligeable contribution. To carry out the $u^\prime$
integration in Eq.~(\ref{int2}) it is convenient to
 add and subtract a term representing
the small $l$ behavior:
\begin{eqnarray}
 && K_2=-2\int\limits_{0}^{\infty}\frac{dl}{l}
\int\limits_{0}^{1} {{du^\prime}\over
{\sqrt{1-u^\prime}}}
\frac{4M_\infty^2}{4M_\infty^2+l^2u^\prime}
\times\nonumber\\
&&\left\{\left[Z_{_{T}}\frac{\omega_{_{T}}^2-l^2}
  {4M_\infty^2-u^\prime(\omega_{_{T}}^2-l^2)}-
  \frac{m_{\hbox{\cmr g}}^2}{4M_\infty^2-m_{\hbox{\cmr g}}^2u^\prime}
\right]-\left[
  \frac{{\rm Re}\,\Pi_{_{T}}(y)}
    {4M_\infty^2-u^\prime{\rm Re}\,\Pi_{_{T}}(y)}
  -\frac{m_{\hbox{\cmr g}}^2}{4M_\infty^2-m_{\hbox{\cmr g}}^2u^\prime}\right]\right\}.
\end{eqnarray}
The in\-te\-gra\-tion o\-ver $u^\prime$ can now be
car\-ried out, 
but the ex\-pli\-cit re\-sults de\-pend on the signs of
$1-4M^2_\infty/\hbox{\rm Re}\,\Pi_{_{T,L}}(\chi)$, 
$1-{4M^2_\infty}/(\omega^2_{_{T,L}}-l^2)$ and
$1-{4M^2_\infty}/m^2_{\hbox{\cmr g}}$, whe\-re 
$\chi=\sqrt{1+4M_\infty^2/l^2}$. 
The quan\-ti\-ty $1-4M^2_\infty/\hbox{\rm Re}\,
\Pi_{_{T,L}}(x)$ is plot\-ted
on Fig.~\ref{figsumrules}, for both the 
transverse and longitudinal
case. Since $\omega^2_{_{T,L}}-l^2=\hbox{\rm Re}\,\Pi_{_{T,L}}(
x\equiv\omega_{_{T,L}}/l)$, and since $\chi$ is varying from $1$ to 
$+\infty$, these curves give us all the signs we need if
we recall also the limits $\lim_{x\to 1^{+}} 
1-4M^2_\infty/\hbox{\rm Re}\,
\Pi_{_{T}}(x)= 1-8M^2_\infty/3m^2_{\hbox{\cmr g}}$ and
$\lim_{x\to 1^{+}} 
1-4M^2_\infty/\hbox{\rm Re}\,
\Pi_{_{L}}(x)= -\infty$.
The case ${3/ 8}<{{M^2_\infty}/{m^2_{\hbox{\cmr g}}}}$
is the most interesting one physically ($e.g.$, an 
$SU(3)$ gauge theory with 
less than 10 light flavors), and fortunately the simplest one. 
We find for the total transverse
contribution the result:
\begin{eqnarray}
&\hbox{\rm Im}\,\Pi^\mu\,_\mu(Q)_{|_{T}}&\approx-{{e^2 g^2}\over{3\pi}}
{T^3\over{q_0}}\int\limits_{0}^{+\infty} {{dl}\over{l}}\;
\left\{
\vphantom{{1\over{1-{{4M^2_\infty}\over{m^2_{\hbox{\cmr g}}}}-z^2}}}
 {{m^2_{\hbox{\cmr g}}}\over
{l^2+m^2_{\hbox{\cmr g}}}}
{1\over \chi}\tanh^{-1}{{1}\over{\chi}}
\right.\nonumber\\
&&+ 
{{4M^2_\infty}\over{l^2+4M^2_\infty}}
{1\over{\sqrt{{{4M^2_\infty}\over{\hbox{\rm Re}\,\Pi_{_{T}}(\chi)}}-1}}}
\tan^{-1}\left({1\over{\sqrt{{{4M^2_\infty}\over
{\hbox{\rm Re}\,\Pi_{_{T}}(\chi)}}-1}}}\right)\nonumber\\
&&
-Z_{_{T}}{{\omega^2_{_{T}}-l^2}\over{\omega^2_{_{T}}}}
{1\over{\sqrt{{{4M^2_\infty}\over{\omega^2_{_{T}}-l^2}}-1}}}
\tan^{-1}\left({1\over{\sqrt{{{4M^2_\infty}\over
{\omega^2_{_{T}}-l^2}}-1}}}\right)\nonumber\\
&&\left.+  
\left(
{{4M^2_\infty}\over{l^2+4M^2_\infty}}
-{{m^2_{\hbox{\cmr g}}}\over
{l^2+m^2_{\hbox{\cmr g}}}}\right)
{1\over{{\sqrt{{{4M^2_\infty}\over{m^2_{\hbox{\cmr g}}}}-1}}}}
\tan^{-1}\left({1\over{{\sqrt{{{4M^2_\infty}\over
{m^2_{\hbox{\cmr g}}}}-1}}}}\right)\right\}.
\end{eqnarray} 
Likewise for the longitudinal case we obtain:
\begin{eqnarray}
&\hbox{\rm Im}\,\Pi^\mu\,_\mu(Q)_{|_{L}}&\approx{{e^2 g^2}\over{3\pi}}
{T^3\over{q_0}}\int\limits_{0}^{+\infty} {{dl}\over{l}}\;
\left\{
\vphantom{{1\over{1-{{4M^2_\infty}\over{m^2_{\hbox{\cmr g}}}}-z^2}}}
 \left({{m^2_{\hbox{\cmr g}}}\over
{l^2+m^2_{\hbox{\cmr g}}}}-{{3m^2_{\hbox{\cmr g}}}\over
{l^2+3m^2_{\hbox{\cmr g}}}}\right)
{1\over \chi}\tanh^{-1}{{1}\over{\chi}}
\right.\nonumber\\
&&+ 
{{4M^2_\infty}\over{l^2+4M^2_\infty}}
{1\over{\sqrt{{{4M^2_\infty}\over{\hbox{\rm Re}\,\Pi_{_{L}}(\chi)}}-1}}}
\tan^{-1}\left({1\over{\sqrt{{{4M^2_\infty}\over
{\hbox{\rm Re}\,\Pi_{_{L}}(\chi)}}-1}}}\right)\nonumber\\
&&
-Z_{_{L}}{{\omega^2_{_{L}}-l^2}\over{\omega^2_{_{L}}}}
{1\over{\sqrt{{{4M^2_\infty}\over{\omega^2_{_{L}}-l^2}}-1}}}
\tan^{-1}\left({1\over{\sqrt{{{4M^2_\infty}\over
{\omega^2_{_{L}}-l^2}}-1}}}\right)\nonumber\\
&&\left.+  
\left(
{{4M^2_\infty}\over{l^2+4M^2_\infty}}
-{{m^2_{\hbox{\cmr g}}}\over
{l^2+m^2_{\hbox{\cmr g}}}}\right)
{1\over{{\sqrt{{{4M^2_\infty}\over{m^2_{\hbox{\cmr g}}}}-1}}}}
\tan^{-1}\left({1\over{{\sqrt{{{4M^2_\infty}\over
{m^2_{\hbox{\cmr g}}}}-1}}}}\right)\right\}.
\end{eqnarray} 

There are two other cases that need to be considered in general:
${1/4}<{{M^2_\infty}/{m^2_{\hbox{\cmr g}}}}<{3/ 8}$ and
${{M^2_\infty}/{m^2_{\hbox{\cmr g}}}}<{1/ 4}$. These two cases
are more intricate because now the required signs depend on
the value of $l$. For this reason, the corresponding expressions will
not be written here. 
There is one limit of
interest, $M_\infty\to 0$, which is included in the latter regime,
but this limit is more conveniently handled at an earlier stage
of the calculation (see Eq.~(\ref{imag})).
One should note that this splitting in three ranges is only a
consequence of the use of sum rules and has no physical
meaning, since it is obvious from 
Eq.~(\ref{imreal2}) that the result is a continuous
function of the ratio
${{M^2_\infty}/{m^2_{\hbox{\cmr g}}}}$.
 The reason for this can be seen from Eq.~(\ref{imreal2}): 
one must extend the $x$ integration range from $[-1,+1]$ to
the whole real axis in order to use sum rules 
and then subtract the contribution of the 
region $]-\infty,-1]\cup[+1,+\infty[$ by Eq.~(\ref{subtraction}). 
While $x\in[-1,+1]$, the
denominator $4M^2_{\infty}+l^2(1-x^2)u'$ remains strictly positive;
but if $x\in]-\infty,-1]\cup[+1,+\infty[$, one can be faced
with a pole in the $u'$ variable, which is dealt with a 
principal part prescription.
 A by-product of this pole 
is that the expression that should be integrated over $l$ depends on
which side of the pole we are.

\subsection{Discussion of the result and asymptotic behavior}

In this paragraph, we study the behavior of 
$\hbox{\rm Im}\,\Pi^\mu\,_\mu(Q)$ in some limiting cases for
${M^2_\infty}$ and ${m^2_{\hbox{\cmr g}}}$. 
We work with the compact form Eq.~(\ref{imag}).
In fact, the quantity of interest is
\begin{equation}
J_{_{T,L}}\equiv \int\limits_{0}^{1}\;
\frac{dx}{x}\;\widetilde{I}_{_{T,L}}(x)\;
\int\limits_{0}^{w^*}\;dw\;
\frac{\sqrt{w/(w+4)}
\hbox{\rm tanh}{}^{-1}\sqrt{w/(w+4)}}{(w+\widetilde{R}_{_{T,L}}(x))^2+
(\widetilde{I}_{_{T,L}}(x))^2},
\label{eqjtl}
\end{equation}
which is dimensionless and is therefore a function of the dimensionless
quantities $l^*/M_\infty$ and $m_{\hbox{\cmr g}}/M_\infty$.
The independence on $l^*$, as well as the infra-red finiteness,  
can be shown to hold using this formula, but since we already did this using
the sum-rule formalism we will not repeat the discussion here. 
We thus assume $J_{_{T,L}}$ to be only a function of 
${{M_\infty}/{m_{\hbox{\cmr g}}}}$ and
study, both analytically and numerically, the limit
of the above expression when the ratio ${{M_\infty}/{m_{\hbox{\cmr g}}}}$
$\to 0$ 
in order to better understand the origin of the singularities.
Of course, the precise value of this ratio is fixed by the number 
of colors and flavors of the studied theory, and is
of order $1$. Despite that fact, this limit enables one to verify 
what is the effect 
of switching the fermion thermal mass off, and gives information
on the nature of the singularities which are regularized by this mass.
Some details of the analysis are given in the appendix. 
We find:
\begin{eqnarray}
&&
J_{_{L}}\;\;\build{M_{\infty}\ll m_{\hbox{\cmr g}}}
\over{\sim}\;\;\ln\left(
{{m_{\hbox{\cmr g}}}\over{M_\infty}}\right)
\nonumber\\
&&
J_{_{T}}\;\;\build{M_{\infty}\ll m_{\hbox{\cmr g}}}
\over{\sim}\;\;\ln^2\left(
{{m_{\hbox{\cmr g}}}\over{M_\infty}}\right).
\label{minfeq0}
\end{eqnarray}
These results are interpreted as follows: the thermal fermion mass 
regularizes the collinear 
divergence, 
and therefore this kind of divergence, common to both the longitudinal 
and
the transverse cases, contributes as one power of 
$\ln(m_{\hbox{\cmr g}}/M_\infty)$. Additionally, in the transverse 
contribution, there is an extra power of that logarithm,
which is a consequence of the potential infrared divergence due to 
the absence of a thermal mass for static transverse gluons. This means
also that this divergence is unexpectedly regularized by the fermion 
thermal mass. In fact, we can see already these features 
in Eq.~(\ref{uinteg}), where the $\tanh^{-1}$ function corresponds to
the collinear logarithmic factor and where the combination 
$\sqrt{-L^2(4M^2_\infty-L^2)}$ (instead of $-L^2$ if $M^2_\infty$
is not taken into account) in the denominator is responsible
for the regularization of potential infrared problems.
  We can also observe in 
that limit the essential difference between a true double
pole and the structure of the integral we denoted by $I_4$ in the
section \ref{enhancement} (indeed, as seen before, $I_4$ is the
integral of interest in our problem, because it is enhanced, and
because it has the dominant part of the numerator associated with
it). If we really had a double pole, like in $I_3$, we would have got
a factor $m^2_{\hbox{\cmr g}}/M^2_\infty$ instead of a logarithm. 
This point will be discussed further when we compare
our results with that of \cite{cleymans,cleyman1,goloviz}. We have checked that
numerical calculations based on Eq.~(\ref{eqjtl}) are in good 
agreement
with the analytical results in the asymptotic region.

 In fact, if the fermion thermal mass is
sufficiently small, there can be a competition between the fermion
mass and an hypothetical magnetic mass, expected at the order 
$m_{\hbox{\cmr mag}}\sim g^2T$. If we introduce this mass 
by the transformation of Eq.~(\ref{magmass1}),
then we obtain:
\begin{itemize}
\item[(i)] If $m_{\hbox{\cmr mag}}\ll M_\infty \ll m_{\hbox{\cmr g}}$:
\begin{equation}
J_{_{T}}\sim \ln^2\left({{m_{\hbox{\cmr g}}}\over{M_\infty}}\right),
\end{equation}
\item[(ii)] If $M_\infty \ll m_{\hbox{\cmr mag}}\ll m_{\hbox{\cmr g}}$:
\begin{equation}
J_{_{T}}\sim \ln\left({{m_{\hbox{\cmr g}}}\over{M_\infty}}\right)\;
\ln\left({{m_{\hbox{\cmr g}}}\over{m_{\hbox{\cmr mag}}}}\right).
\end{equation}
\end{itemize}
This competition between the magnetic mass and the fermion mass has
been studied numerically, which gives the curves of Fig.~(\ref{figmag}),
where are plotted the value of $J_{_{T}}$ as a function
of $\log({m_{\hbox{\cmr mag}}/m_{\hbox{\cmr g}}})$, for various
fixed values of the ratio $M_\infty/m_{\hbox{\cmr g}}$. We observe 
that the effect of the magnetic mass becomes non negligeable 
as far as $m_{\hbox{\cmr mag}}\geq 0.1\,M_\infty$. When the
magnetic mass is small enough, the flattening of the curves denotes
the independence of the result on $m_{\hbox{\cmr mag}}$.

\section{Comparison with the semi-classical results}
The purpose of this section is to compare the results provided
by the thermal field theory techniques with those obtained via
semi-classical methods.

\subsection{Structure of the emission rate}
In the semi-classical approach, one finds 
\cite{cleymans,cleyman1,goloviz,land-lif} that
the emission rate of soft photons 
in a scattering process 
can be factorized
as the product of two terms: the intensity of emission which is
the square of an electromagnetic current, and the cross section
for the same scattering process without photon emission. 
Our purpose is to make a precise connection between such a result and
our result Eq.~(\ref{sum}).

Let us begin by defining the amplitude corresponding to the
scattering process without any photon emission, i.e. corresponding
to the diagrams of Fig.~\ref{figampl}-(d),(e) where the
photon line is suppressed (written here for the scattering of
a quark on another quark, but it is very simple to write it
 also  for the scattering on a gluon; moreover, like 
in Eq.~(\ref{sum}), we don't write the color and group factors):
\begin{eqnarray}
&i{\cal M} (P_1,P_2;P'_1,P'_2;L)&\equiv 
-{\bar u}(P'_1)(-ig\gamma^\mu)u(P_1)\;
{\bar u}(P'_2)(-ig\gamma^\nu)u(P_2) \nonumber\\
&&
\times\sum\limits_{a=T,L} P^a_{\mu\nu}(L)
{}^{*}\Delta^a(L),
\end{eqnarray}
where $P_1$, $P_2$ are the incoming momenta, 
$P'_1$, $P'_2$ are the outgoing momenta, and $L$ is the
transferred momenta.
Of course, only three of these five momenta are independent, and
appropriate Dirac delta functions will be needed to enforce the 
momentum conservation. The amplitude squared is simply given by:
\begin{equation}
|{\cal M}|^2=g^4\; 
{\rm Tr}(\gamma^\mu\slP_1\gamma^\sigma\slP'_1)\;
{\rm Tr}(\gamma^\nu\slP_2\gamma^\rho\slP'_2)
\sum\limits_{a,b=T,L} P^a_{\mu\nu}(L)P^b_{\rho\sigma}(L)\,
{}^{*}\Delta^a(L) 
\left[{}^{*}\Delta^b(L)\right]^{*}.
\label{amplitude}
\end{equation}

We are now going to rewrite  Eq.~\ref{sum} into a form exhibiting 
the decomposition mentioned above.
In order to make the connection with Eq.~\ref{sum}, we
first notice:
\begin{equation}
4(R_\rho R_\sigma+P_\rho P_\sigma)\approx {\rm Tr}(\gamma^\rho
(\slR+\slL)\gamma^\sigma \slP),
\label{approxtrace} 
\end{equation}
and 
\begin{equation}
{{4 e^2 L^2}\over{(R^2-M^2_\infty)((P+L)^2-M^2_\infty)}}\approx
-e^2\sum\limits_{\hbox{\cmr pol.}\ \varepsilon}\left(
{{P\cdot\varepsilon}\over{P\cdot Q}}-
{{(R+L)\cdot\varepsilon}\over{(R+L)\cdot Q}}
\right)^2,
\label{approxcurrent}
\end{equation}
where the sum runs over the polarizations of the emitted photon.
We recognize the standard electromagnetic current which appears
in such a soft emission \cite{land-lif}. It is worth recalling that we 
used the fact that the relevant values of $u$ are of order 
$u^*\ll 1$.
We need also the hard thermal loop contribution to the 
imaginary part of the polarization tensor 
of the gluon:
\begin{eqnarray}
\hbox{\rm Im}\,\Pi^{\mu\nu}_{\hbox{\cmr gluon}}(L)={{g^2}\over{2}}
\int{{d^4K}\over{(2\pi)^4}}&&
\left[n_{_{F}}(k_0+l_0)-n_{_{F}}(k_0)\right]\times\nonumber\\
&&2\pi\delta(K^2)\; 2\pi\delta((K+L)^2)\;
{\rm Tr}(\gamma^\mu\slK\gamma^\nu(\slK+\slL)),
\end{eqnarray}
where we considered only the quark loop contribution. Of course, the
gluon loop would be related to the scattering of the quark on a gluon,
which has not been written.
Moreover, like in Eq.~(\ref{sum}) and Eq.~(\ref{amplitude}),
 we have not written the color and
group factors.
Concerning the statistical weights of Eq.~(\ref{sum}),
a straightforward calculation gives
 (the $n_{_{B}}(q_0)$ factor should be added in order to
reconstruct the emission rate defined in 
Eq.~(\ref{rate}), and the other statistical factors have been taken
at an earlier stage of the calculation, for example in Eqs.~(\ref{self}) 
and (\ref{vertex}), before any approximation):
\begin{eqnarray}
&\displaystyle{n_{_{B}}(q_0)
\left[n_{_{B}}(l_0)+n_{_{F}}(r_0+l_0)\right]}&
\left[n_{_{F}}(r_0)-n_{_{F}}(p_0)\right]
\left[n_{_{F}}(k_0+l_0)-n_{_{F}}(k_0)\right]\nonumber\\
&&=n_{_{F}}(r_0+l_0)\left[1-n_{_{F}}(p_0)\right]n_{_{F}}(k_0)
\left[1-n_{_{F}}(k_0+l_0)\right].
\end{eqnarray}
We need also the following identity:
\begin{eqnarray}
&{\rm Im}\,\Pi^{\nu\rho}_{\hbox{\cmr gluon}}(L)&\sum\limits_{a,b=T,L}
P^a_{\mu\nu}(L)P^b_{\rho\sigma}(L)\,
{}^{*}\Delta^a(L) 
\left[{}^{*}\Delta^b(L)\right]^{*}\nonumber\\
&&=-\sum\limits_{a,b,c=T,L}{\rm Im}\,\Pi_{c}(L) P^{\nu\rho}_{c}(L) 
P^a_{\mu\nu}(L)P^b_{\rho\sigma}(L)\,
{}^{*}\Delta^a(L) 
\left[{}^{*}\Delta^b(L)\right]^{*}\nonumber\\
&&=-\sum\limits_{a=T,L}{\rm Im}\,\Pi_{a}(L) 
|{}^*\Delta^a(L)|^2 P^a_{\mu\sigma}(L)\nonumber\\
&&=-{1\over 2} \sum\limits_{a=T,L} \rho_a(L) P^a_{\mu\sigma}(L),
\end{eqnarray}
the last equality being valid in the region where $L^2<0$.

Given all these relations, it is straightforward to
rewrite our result for the
emission rate 
(see Eqs.~(\ref{rate})
and (\ref{sum})) per unit time and unit volume as:
\begin{eqnarray}
\openup 15mm
&\displaystyle{{{dN}\over{d^4x}}
\approx {{d^3q}\over{(2\pi)^3 2q_0}}}&
\int\prod\limits_{i=1,2}{{d^4P_i}\over{(2\pi)^4}}
\prod\limits_{i=1,2}{{d^4P'_i}\over{(2\pi)^4}}
(2\pi)^4 \delta(P_1+P_2-P'_1-P'_2-Q)
\nonumber\\
&&\times 2\pi n_{_{F}}(p_1^0)\delta(P_1^2-M^2_\infty)\;
2\pi \left[1-n_{_{F}}(p^{'0}_1)\right]\delta(P_1^{'2}-M^2_\infty)\nonumber\\
&&\times 2\pi n_{_{F}}(p_2^0)\delta(P_2^2)\;
2\pi \left[1-n_{_{F}}(p^{'0}_2)\right]\delta(P_2^{'2})\nonumber\\
&&\times |{\cal M}|^2(P_1,P_2;P'_1+Q,P'_2;L\equiv P'_2-P_2)\nonumber\\
&&\times e^2\sum\limits_{\hbox{\cmr pol.}\ \varepsilon}\left(
{{P_1\cdot\varepsilon}\over{P_1\cdot Q}}-
{{P'_1\cdot\varepsilon}\over{P'_1\cdot Q}}
\right)^2,
\label{factorization}
\end{eqnarray}
where $|{\cal M}|^2$ is given in Eq.~(\ref{amplitude}).
This expression shows clearly that thermal field theory 
calculation leads also
to a separation in two factors: the amplitude squared
of the similar scattering
process without photon emission, and a factor which is nothing but the
square of an electromagnetic current and is specific 
to the photon emission. These two factors are then integrated 
over the momenta of the unobserved particles, with appropriate
statistical weights. Because of this high similarity with the
semi-classical expressions, this structure can be seen as
another evidence to say that the enhanced terms we have
exhibited are actually the relevant contributions to
photon production by a plasma.

This decomposition is valid only when the photon momentum $Q$ is
 negligeable
in front of the quark momentum, so that the amplitude that 
appears in Eq.~(\ref{factorization}) is almost unperturbed 
by the momentum $Q$.
It is worth saying that we used several times the fact that $u^*\ll 1$
to give that structure to Eq.~(\ref{sum}). It means that this kind
of factorization will not be valid in the case where the virtuality 
is important $Q^2/q_0^2\sim 1$. Moreover, compared to the standard
semi-classical treatments \cite{cleymans,cleyman1,goloviz}, we
see that this structure remains valid even if we take into
account the transverse exchange contribution, which has
been shown in the previous section to be of the same order of
magnitude as the longitudinal one.

\subsection{Factorization limit}

Nevertheless, despite that nice structure, the phase space over which 
the integration has to be performed is common to the two factors.
Indeed, these two factors share some variables, like the
momentum $L$ exchanged during the scattering. Therefore, from
a technical point of view, the factorization is not really complete.

The integral over the  angles $\theta$ and $\phi$ of
the photon line affect only the
factor related to the photon emission, and is easily obtained via
Eqs.~(\ref{approxcurrent}), (\ref{deno1}), (\ref{deno2}) and 
(\ref{uinteg}):
\begin{equation}
\int\limits_{0}^{2\pi}d\phi
\int\limits_{0}^{2} du 
\sum\limits_{\hbox{\cmr pol.}\ \varepsilon}\left(
{{P_1\cdot\varepsilon}\over{P_1\cdot Q}}-
{{P'_1\cdot\varepsilon}\over{P'_1\cdot Q}}
\right)^2\approx {{16\pi}\over{q_0^2}}
\sqrt{{-L^2}\over{4M^2_\infty-L^2}}\tanh^{-1} 
\sqrt{{-L^2}\over{4M^2_\infty-L^2}}.
\end{equation}
As already said, the $\tanh^{-1}$ factor is a remnant of the collinear
singularity (when $M_\infty\to 0$, this factor diverges
logarithmically). 

Then, we see that this emission 
factor still depends on the variable $L^2$, so that the 
apparent factorization provided by 
Eq.~\ref{factorization} is not a true factorization. 
We can look at the two limiting cases:
\begin{itemize} 
\item[(i)] If $L\ll M_\infty$, we have:
\begin{equation}
\int\limits_{0}^{2\pi}d\phi
\int\limits_{0}^{2} du 
\sum\limits_{\hbox{\cmr pol.}\ \varepsilon}\left(
{{P_1\cdot\varepsilon}\over{P_1\cdot Q}}-
{{P'_1\cdot\varepsilon}\over{P'_1\cdot Q}}
\right)^2\approx {{-4\pi L^2}\over{q_0^2 M^2_\infty}}.
\label{smallL}
\end{equation}
Therefore, the coupling between $L$ and $M_\infty$ in 
the emission factor
will be useful at small $L$ to regularize potentially
dangerous terms in the infrared sector. This is the reason why 
the transverse gluon exchange is infrared finite despite
the lack of thermal mass for static transverse modes.
\item[(ii)] If $L\gg M_\infty$, we have:
\begin{equation}
\int\limits_{0}^{2\pi}d\phi
\int\limits_{0}^{2} du 
\sum\limits_{\hbox{\cmr pol.}\ \varepsilon}\left(
{{P_1\cdot\varepsilon}\over{P_1\cdot Q}}-
{{P'_1\cdot\varepsilon}\over{P'_1\cdot Q}}
\right)^2\approx 
{{8\pi}\over{q_0^2}} \ln\left({{-L^2}\over{4M^2_\infty}}\right).
\end{equation}
This limit is known as the ``factorization limit" \cite{peigne}. 
Indeed, in that limit,
the $L$ dependence of the emission factor is only logarithmic and is
therefore negligeable, so that the only relevant $L$ dependence is
in the scattering amplitude. 

This limit is valid for the longitudinal
contribution when $m_{\hbox{\cmr g}}\gg M_\infty$, because
in that case the relevant infrared regulator is $m_{\hbox{\cmr g}}$
(which is the thermal mass of static longitudinal modes) and
the relevant values of $L$ are also of
order $L\sim m_{\hbox{\cmr g}}\gg M_\infty$. In that limit,
the longitudinal contribution becomes proportional to 
$\ln(m^2_{\hbox{\cmr g}}/M^2_\infty)$.

Almost the same interpretation can be given for the transverse 
contribution: we get one factor $\ln(m^2_{\hbox{\cmr g}}/M^2_\infty)$
from the $\tanh^{-1}$ when $m_{\hbox{\cmr g}}\gg M_\infty$,
whereas the second power of that logarithm comes from the
singular infrared behavior of the scattering amplitude. 
In that case, the ``factorization limit" cannot be used for all
the values of $L$ since values of $L$ of order $M_\infty$ are explored 
due to the lack of transverse static thermal mass.
\end{itemize}

In conclusion, we can say that the expression Eq.~(\ref{factorization}),
despite its high similarity with semi-classical ones, should
be interpreted with care concerning the possibility of
``factorization". Moreover, this lack of factorization at
small $L$ is closely related to the possibility of getting a finite
transverse contribution.

\subsection{Order of magnitude}

After having seen that our expression for the photon
production rate, derived from thermal field theory,
is similar to the expressions usually 
encountered in semi-classical methods, it remains to compare the
order of magnitude of the rate given by the two methods since
the calculations are performed in very different ways.

From Eq.~(\ref{rate}) and the discussion in the previous section,  the
differential rate for real photon production, integrated over angular
variables, takes the form
\be
{\frac{d N} {dq_0 d^4x}} \approx \frac{e^2 g^2 N_{_{C}} C_{_{F}}}{6\pi^4}
  \frac{T^4}{q_0} (-1)_{_{L}}J_{_{T,L}}
  \left({{m_{\hbox{\cmr g}}}\over{M_\infty}}\right)
\label{oureq}
\ee
where one recognizes the typical $1 / q_0$ bremsstrahlung spectrum. 
It is
instructive to compare our result with that in 
\cite{cleymans,cleyman1,goloviz} in the case the LPM effect is turned off.
It is found there, neglecting numerical constants, that
in the semi-classical approximation
\be
{\frac{d N} {dq_0d^4x}} \approx e^2 g^4 
\frac{T^6}{ q_0 M^2_\infty}
\ln \left( {T^2 \over m_{\hbox{\cmr g}}^2} \right).
\label{theireq}
\ee
The structure of this
equation can be understood as follows: the $ T^2 / M_\infty^2$ piece
is arising from the collinear radiation of the photon by the hard quark,
while the factor $\ln (T^2 /  m_{\hbox{\cmr g}}^2)$ reflects the
divergence of the quark scattering cross section in the plasma,
screened by the Debye mass $m_{\hbox{\cmr g}}$. 
Eqs. (\ref{oureq}) and 
(\ref{theireq}) are in qualitative agreement since 
$M^2_\infty \sim g^2 T^2$
and both rates are of order $e^2 g^2 T^4 / q_0$.
In fact the enhancement mechanism discussed above is realized in 
\cite{cleymans,cleyman1,goloviz} by the factor $T^2 / M^2_\infty$.
We interpret this pole in $M^2_\infty$ as a use of the
approximation of Eq.~(\ref{smallL}), extended to the
region where $L\sim T$ in order to get the logarithmic factor
$\ln(T^2/M^2_{\hbox{\cmr g}})$.
This leads to a
very different limiting behavior when $M_\infty \to 0$ since in our approach
we recover a logarithmic singularity while in the semi-classical approach
one obtains a pole in $M_\infty$ \cite{goloviz}.
Another difference between Eq.~(\ref{oureq}) and the results of Cleymans
$et$ $al.$ is that the transverse gluon exchange gives a finite
contribution which is equally important as that of the screened longitudinal
one considered in \cite{cleymans,cleyman1,goloviz}. As a consequence, the 
use of the static approximation to describe the interaction of the quark 
in the plasma is not justified in thermal field theory.

\section{The slightly virtual case: $\lowercase{g}^2\le 
Q^2/\lowercase{q}_0^2\ll 1$}

We now consider the case (ii) of Sec.~\ref{cases}
where the virtuality $Q^2$ is
non negligeable but remains small. This
can be treated in a simple way thanks to the introduction of an
effective fermion mass.
\subsection{Expression of $\mbox{\rm Im}\Pi^\mu\,_\mu$}
As already mentioned, in this case we still have a clear hierarchy
in the various powers of $Q^2/q_0^2 \ll1$. The consequence of this
hierarchy is that the dominant term of the numerator
remains the same as in 
 Eqs.~(\ref{sum}), and that it can still be approximated 
by Eq.~(\ref{numerator}). In what concerns the denominators, we saw in
Eqs.~(\ref{deno1}),~(\ref{deno2}) that their expression remains 
unchanged provided one replaces the fermion thermal mass $M_\infty$ by
an effective mass taking into account the effect of the virtuality
of the emitted photon $M^2_{\hbox{\cmr eff}}\equiv M^2_\infty+
Q^2r^2/q_0^2$. Even before writing the expression of the
imaginary part of the photon self-energy, we can notice that
the positive virtuality $Q^2>0$ has the consequence of increasing the
effect of the fermion thermal mass. 
Another point to be looked at is the influence of the constraint
$\cos\theta^\prime\in[-1,1]$ on the phase space. By examining the
denominators entering in the rate, we easily see that the relevant
values of $u$ in the integration over $u$ are of order of
$u^*\sim M^2_{\hbox{\cmr eff}}/r^2$. As far as $Q^2/q_0^2\ll 1$,
this collinear regulator is much smaller than $1$.
 Therefore, we can still
approximate $\cos\theta^\prime \approx l_0/l$. Therefore, the
effect of the constraint $\cos\theta^\prime\in[-1,1]$ is still
to confine $L^2$ to negative values. Finally, in complete analogy with
Eq.~(\ref{imreal}), the imaginary part  of the self-energy of a virtual photon
reads:
\begin{eqnarray}
&\hbox{\rm Im}\,\Pi^\mu\,_\mu(Q)&\approx
(-1)_{_{T}}{{e^2 g^2}\over{8\pi^4}}
{1\over{q_0}}\int\limits_{r^*}^{\infty}dr n_{_{F}}^\prime(r)
\int\limits_{0}^{l^*}l^4 dl \int_{-1}^{+1}{dx}\,n_{_{B}}(lx)\, 
\rho_{_{T,L}}(l,lx)(1-x^2)^2\nonumber\\
&&\times\int\limits_{0}^{2} {{du}\over
{\left[u+{{M_{\hbox{\cmr eff}}^2}\over{2r^2}}
\right]\left[\left(u+{{M_{\hbox{\cmr eff}}^2}\over
{2r^2}}+{{l^2(x^2-1)}\over{2r^2}}
\right)^2+{{l^2(1-x^2)M_{\hbox{\cmr eff}}^2}\over{r^4}}\right]^{1/2}}},
\label{imvirtual}
\end{eqnarray}
the notations being the same as in Eq.~(\ref{imreal}).

\subsection{Reduction of the expression}
The simplification of this imaginary part can be performed by the
same tools as in Sec.~\ref{reduction}. In particular, since
the relevant values of the angular variable $u$ are much smaller
than one, the same approximations are valid. Despite that
similarity, an essential difference lies in the fact that the
integration over $r$ cannot be factorized since the effective fermion
mass now depends on $r$. If one introduces the
dimensionless variables:
\begin{eqnarray}
&&v\equiv {{r}\over{T}}\nonumber\\
&&w\equiv {{-L^2}\over{M^2_{\hbox{\cmr eff}}(v)}}\nonumber\\
&&\widetilde{I}_{_{T,L}}(x,v)\equiv {{\hbox{\rm Im}\,
\Pi_{_{T,L}}(x)}\over{M^2_{\hbox{\cmr eff}}(v)}}\nonumber\\
&&\widetilde{R}_{_{T,L}}(x,v)\equiv {{\hbox{\rm Re}\,
\Pi_{_{T,L}}(x)}\over{M^2_{\hbox{\cmr eff}}(v)}},
\end{eqnarray}
Eq.~(\ref{imvirtual}) takes the form
\begin{eqnarray}
& & \hbox{\rm Im}\,\Pi^\mu\,_\mu(Q)\approx(-1)_{_{T}}
  \frac{2 e^2 g^2 N_{_{C}} C_{_{F}}}{\pi^4}
  \frac{T^3}{q_0}\int\limits_{r^*/T}^{+\infty} 
dv\,v^2\,{1\over{e^v+1}}\left[1-{1\over{e^v+1}}\right]\nonumber\\
& & \qquad\times\int\limits_{0}^{1}\;
\frac{dx}{x}\;\widetilde{I}_{_{T,L}}(x,v)\int\limits_{0}^{w^*(v)}\;dw\;
\frac{\sqrt{w/(w+4)}
\hbox{\rm tanh}{}^{-1}\sqrt{w/(w+4)}}{(w+\widetilde{R}_{_{T,L}}(x,v))^2+
(\widetilde{I}_{_{T,L}}(x,v))^2}.
\label{imagvirtual}
\end{eqnarray}
where, compared to Eq.~(\ref{imag}), the integration over $v$ cannot be
factorized because 
of the $v$ dependence contained in the quantities $w^*$,
$\widetilde{R}_{_{T,L}}$ and $\widetilde{I}_{_{T,L}}$.

\subsection{Discussion of the result}
When the virtuality $Q^2>0$ is switched on, the quantity which is
worth studying is now the integral:
\begin{eqnarray}
&\displaystyle{J_{_{T,L}}\equiv{{6}\over{\pi^2}}}& 
\int\limits_{r^*/T}^{+\infty}dv\, v^2\,
{1\over{e^v+1}}\left[1-{1\over{e^v+1}}\right]\;
\int\limits_{0}^{1}\;
\frac{dx}{x}\;\widetilde{I}_{_{T,L}}(x,v)\;\nonumber\\
&&\times\int\limits_{0}^{w^*(v)}\;dw\;
\frac{\sqrt{w/(w+4)}
\hbox{\rm tanh}{}^{-1}\sqrt{w/(w+4)}}{(w+\widetilde{R}_{_{T,L}}(x,v))^2+
(\widetilde{I}_{_{T,L}}(x,v))^2},
\label{Jvirtual}
\end{eqnarray}
where the explicit prefactor $6/\pi^2$ is introduced to compare
this definition for $J_{_{T,L}}$ with the previous
one in the limit where $Q^2=0$ and where the integral over $v$
can be factorized.

{\it A priori}, the functions $J_{_{T,L}}$ depend of the
four dimensionless quantities $r^*/T$, $l^*/M_\infty$, 
$M_\infty/m_{\hbox{\cmr g}}$ and $Q^2T^2/q_0^2 m^2_{\hbox{\cmr g}}$.
Nevertheless, since the weight function $p(v)\equiv 
v^2 e^v/(e^v+1)^2$ is
peaked around $v\sim 1$, the small values of $v$ will be negligeable
in the result, which means that in fact this result does not
depend on the value of the cut-off $r^*$, which thus can be taken to zero.

Another question which can be answered very simply by using the
study done for the production of real photons
concerns the absence of IR divergences when the emitted
photon is virtual. Indeed, we saw earlier that the IR finiteness
was in fact due to the presence of a fermion thermal mass. 
In the present case, the effect of this fermion mass is
enhanced by the presence of a positive virtuality $Q^2>0$, since
$M_{\hbox{\cmr eff}}> M_\infty$. Therefore, it is obvious that the
production rate of virtual photons will be IR safe, even for
the contribution of the transverse gluon, and without the
need of a magnetic mass.

To end with generalities, one can notice that the
functions $J_{_{T,L}}$ are decreasing functions of the virtuality
$Q^2$. Indeed, if $Q^2$ increases, the effective mass 
$M_{\hbox{\cmr eff}}(r)$ increases uniformly in $r$. Then, according
to the study performed in the case of real photons, the functions
$J_{_{T,L}}$ should decrease.

Concerning the dependence on the UV cut-off $l^*$, one can 
evaluate the contribution of the region $l\in [l^*,+\infty[$ as
being:
\begin{eqnarray}
&&J^\prime_{_{T}}\approx {{9}\over{2\pi}}\left({{m_{\hbox{\cmr g}}}
\over{l^*}}\right)^2\;
\int\limits_{0}^{+\infty}dv\,p(v)\,
\ln\left({{l^*}\over{M_{\hbox{\cmr eff}}(v)}}\right)\nonumber\\
&&J^\prime_{_{L}}\approx -{{9}\over{\pi}}\left({{m_{\hbox{\cmr g}}}
\over{l^*}}\right)^2\;
\int\limits_{0}^{+\infty}dv\,p(v)\,
\ln\left({{l^*}\over{M_{\hbox{\cmr eff}}(v)}}\right).
\label{hardl}
\end{eqnarray}
Neglecting the logarithms, the contribution of the hard region 
$[l^*,+\infty[$ is of order $(m_{\hbox{\cmr g}}/l^*)^2$, and 
must be compared to $J_{_{T,L}}$ in order to
know if we can make $l^*=+\infty$. 
If we recall that the values which contribute dominantly
in the integral over $v$ are of order $1$,
 we see that the influence of the 
virtuality $Q^2>0$ on $J_{_{T,L}}$
is controlled by the comparison of 
$M^2_\infty$ and $Q^2T^2/ q_0^2$ (see Eq.~(\ref{masseff})). The results 
are as follows:
\begin{itemize}
\item[(i)] If $Q^2/q_0^2\ll M^2_\infty/T^2\sim g^2$, the virtuality
$Q^2$ has no effect and can be neglected: this case corresponds to
the previous section. The functions $J_{_{T,L}}$ are of order $1$,
and the contribution of $[l^*,+\infty[$ is negligeable as far as
$l^*\gg m_{\hbox{\cmr g}}$. It means that the hard region around
$l\sim T$ does not contribute.

\item[(ii)] If $Q^2/q_0^2\sim g^2$, we are in the intermediate
region where the virtuality of the radiated photon and the fermion
thermal mass have comparable effects. The functions $J_{_{T,L}}$ are 
still of order $1$, and the conclusion of the previous case remains 
unchanged.

\item[(iii)]  If $g^2 \ll Q^2/q_0^2$, the effect of the virtuality 
becomes much
more important than the effect of the fermion thermal mass $M_\infty$,
so that $M^2_{\hbox{\cmr eff}}\approx Q^2r^2/q_0^2$.
Moreover, in that case, we have $m_{\hbox{\cmr g}}\ll 
M_{\hbox{\cmr eff}}(v\sim 1)$. By using the same tools as in 
the previous section, one obtains easily the simplified expression:
\begin{equation}
J_{_{T,L}}(l^*\to\infty)\approx{3\over{2\pi^2}}\int\limits_{0}^{+\infty}
dvv^2{1\over{e^v+1}}\left[1-{1\over{e^v+1}}\right]
\ln\left({{M^2_{\hbox{\cmr eff}}(v)}\over{m^2_{\hbox{\cmr g}}}}\right)
\int\limits_{0}^{1}{{dx}\over{x}} \widetilde{I}_{_{T,L}}(x,v)
\end{equation}
and, separating the transverse and longitudinal terms, one gets:
\begin{eqnarray}
&&J_{_{T}}(l^*\to\infty)\approx{{3}\over{8\pi}}\,
{{q_0^2 m^2_{\hbox{\cmr g}}}\over{Q^2T^2}}
\ln\left({{Q^2T^2}\over{q_0^2 m^2_{\hbox{\cmr g}}}}
\right)\nonumber\\
&&J_{_{L}}(l^*\to\infty)\approx -2J_{_{T}}(l^*\to\infty),
\label{bigQ}
\end{eqnarray}
where the $v$ dependence is neglected inside the logarithm.
We have checked that this approximation is in agreement with 
numerical estimates of Eq.~(\ref{Jvirtual}).
We are now in position to compare the contribution from $[l^*,+\infty[$
with that of $[0,l^*]$. To make the discussion more
precise, let us assume that $Q^2/q_0^2=g^\alpha$ and $l^*=g^\beta T$,
with $\alpha,\beta >0$. Then, neglecting the logarithmic factors
in our comparison, we have:
\begin{eqnarray}
&&J'_{_{T,L}}\ll J_{_{T,L}}\nonumber\\
&\Longleftrightarrow&J'_{_{T,L}}\ll J_{_{T,L}}(l^*\to\infty)\nonumber\\
&\Longleftrightarrow&g^\alpha\ll g^{2\beta}\nonumber\\
&\Longleftrightarrow&\beta < \alpha/2\qquad
\hbox{\rm since }\quad g\ll 1.
\end{eqnarray}
This means that for fixed $Q^2/q_0^2=g^\alpha\ll 1$, 
the dependence on $l^*$
disappears as long as $l^*>g^\beta T$ with $0<\beta<\alpha/2$, and so
 the contribution
to the result of the region beyond such a $l^*$ is negligeable.
Therefore, in the region which contributes dominantly to the result,
we still have $l\leq g^\beta T\ll T$, since $\beta > 0$. This
means that our kinematical approximations are safe.
Of course, this works since we had $Q^2/q_0^2 \ll 1$ so that 
$\alpha$ is strictly positive. When the ratio $Q^2/q_0^2$ 
approaches the value $1$, the negligeable region around $l\sim T$
becomes smaller and smaller, so that the precision of our
approximations decreases. Another way to say the same thing is that
 given a fixed value of $g\ll 1$ and given 
an accuracy one wants to ensure, there exists an $\alpha>0$
so that $Q^2/q_0^2$ must remain lower than $g^\alpha$.
In order to reach values of $Q^2/q_0^2$ closer to $1$, one
has to relax the constraint on the accuracy or to consider
a smaller $g$.
We emphasize that, although we have
introduced here fractional powers of the coupling constant $g$
in order to quantify when our approximations for evaluating
the bremsstrahlung diagrams are valid, this does not mean that 
the perturbation expansion, in integer powers of $g$, breaks-down;
nevertheless, before calculating next-to-leading order
diagrams, one should improve the approximations for the
leading ones.
\end{itemize}

A general consequence of
this analysis is the fact that the relevant scales
in our problem are not only $gT$ and $T$ like in the standard 
HTL framework, but also the scale
of the effective mass $M_{\hbox{\cmr eff}}(v\sim 1)\approx 
M^2_\infty+Q^2T^2/q_0^2$ which
is intermediate between $gT$ and $T$. This is a consequence of the
fact
 that the regularization of all our divergences, including the
infrared one,
is done by this effective mass rather than by the resummation
of the HTLs in the gluon propagator.

The case $Q^2/q_0^2\sim 1$
does not fit in these approximations since in that case $\alpha$ is
of order zero, so that $\beta$ is also of order zero and
even the hard
values of $l$ contributes to the result.

To summarize this section, the functions $J_{_{T,L}}$ have been
plotted as a function of $\log(Q^2/q_0^2)$ on Fig.~\ref{figQ} 
over the whole range of admissible values of $Q^2$. We observe
a clear flattening of the curves for small enough $Q^2/q_0^2$,
indicating that the dependence on $Q^2$ becomes negligeable.
In the opposite direction, we see that the enhancement of the rate
decreases if $Q^2\to q_0^2$. The enhancement (ratio between the values
of $J_{_{T,L}}$ at $Q^2/q_0^2=0$ and at $Q^2/q_0^2\sim 1$) 
decreases also if the coupling constant $g$ increases.
We note also that the longitudinal
and transverse contributions are almost equal in magnitude.

\subsection{Reduction using sum rules}
Again, it is possible to reduce Eq.~(\ref{imvirtual}) 
 by performing exactly the integral over $x$ thanks to sum rules.
Since now we cannot factorize the integral over $r$, the result
given by sum rules is a two-dimensional integral. The techniques to
achieve that are exactly the same as in Sec.~\ref{sumrules}, and
therefore will not be reproduced here. We still need to
distinguish three cases. For example, for the phenomenologically
relevant case, 
${3/ 8}<{{M^2_\infty}/{m^2_{\hbox{\cmr g}}}}$, the quantities
$1-4M^2_{\hbox{\cmr eff}}(r)/\hbox{\rm Re}\,\Pi_{_{T,L}}(\chi)$, 
$1-{4M^2_{\hbox{\cmr eff}}(r)}/(\omega^2_{_{T,L}}-l^2)$ and
$1-{4M^2_{\hbox{\cmr eff}}(r)}/m^2_{\hbox{\cmr g}}$ are negative for every
value of $r$. The integration over $u'$ gives 
in the transverse case:
\begin{eqnarray}
&\hbox{\rm Im}\,\Pi^\mu\,_\mu(Q)_{|_{T}}&\approx-{{2 e^2 g^2}\over{\pi^3}}
{T^3\over{q_0}}
\int\limits_{0}^{+\infty}dv\, v^2\,
{1\over{e^v+1}}\left[1-{1\over{e^v+1}}\right]
\int\limits_{0}^{+\infty} {{dl}\over{l}}\;
\left\{
\vphantom{{1\over{1-{{4M^2_{\hbox{\cmr eff}}(v)}\over
{m^2_{\hbox{\cmr g}}}}-z^2}}}
 {{m^2_{\hbox{\cmr g}}}\over
{l^2+m^2_{\hbox{\cmr g}}}}
{1\over \chi}\tanh^{-1}{{1}\over{\chi}}
\right.\nonumber\\
&&+ 
{{4M^2_{\hbox{\cmr eff}}(v)}\over{l^2+4M^2_{\hbox{\cmr eff}}(v)}}
{1\over{\sqrt{{{4M^2_{\hbox{\cmr eff}}(v)}\over{\hbox{\rm Re}\,
\Pi_{_{T}}(\chi)}}-1}}}
\tan^{-1}\left({1\over{\sqrt{{{4M^2_{\hbox{\cmr eff}}(v)}\over
{\hbox{\rm Re}\,\Pi_{_{T}}(\chi)}}-1}}}\right)\nonumber\\
&&
-Z_{_{T}}{{\omega^2_{_{T}}-l^2}\over{\omega^2_{_{T}}}}
{1\over{\sqrt{{{4M^2_{\hbox{\cmr eff}}(v)}\over{\omega^2_{_{T}}-l^2}}-1}}}
\tan^{-1}\left({1\over{\sqrt{{{4M^2_{\hbox{\cmr eff}}(v)}\over
{\omega^2_{_{T}}-l^2}}-1}}}\right)\nonumber\\
&&\left.+  
\left(
{{4M^2_{\hbox{\cmr eff}}(v)}\over{l^2+4M^2_{\hbox{\cmr eff}}(v)}}
-{{m^2_{\hbox{\cmr g}}}\over
{l^2+m^2_{\hbox{\cmr g}}}}\right)
{1\over{{\sqrt{{{4M^2_{\hbox{\cmr eff}}(v)}\over
{m^2_{\hbox{\cmr g}}}}-1}}}}
\tan^{-1}\left({1\over{{\sqrt{{{4M^2_{\hbox{\cmr eff}}(v)}\over
{m^2_{\hbox{\cmr g}}}}-1}}}}\right)\right\}.
\end{eqnarray} 
%
%
%
%
For the other cases ${1/4}<{{M^2_\infty}/{m^2_{\hbox{\cmr g}}}}<{3/ 8}$
and ${{M^2_\infty}/{m^2_{\hbox{\cmr g}}}}<{1/ 4}$ similar calculations
can be carried out but they are much more complex since the
signs which control the expressions depend on both $v$ and $l$.

\section{The case: $Q^2/\lowercase{q}_0^2\sim 1$}
This case will not be studied in detail in this paper; we 
discuss only why it is very different from the previous ones,
and what is the expected order of magnitude of the production rate.
This case differs from the previous two in the following respects:
\begin{itemize}
\item[(i)] Since $Q^2/q_0^2\sim1$, all powers of this ratio should
now be kept in the expansion of the quantities we need.
\item[(ii)] The relevant values of the angular variable $u$
are now of order $u^*\sim 1$. Therefore, the collinear approximation
(which means keeping only the  term in $u^0$ in the numerator)
cannot be used. For the same reason, the approximation
$\cos\theta'\approx l_0/l$ is no longer valid. This means that
the contribution of the domain $L^2>0$ is as important as the 
contribution of the Landau damping $L^2<0$. Physically, this
corresponds to processes involving the Compton effect (a thermalized
on-shell gluon is absorbed by the moving quark, which emits a photon).
\item[(iii)] The scattering of the quark is now sensitive to 
parton exchanges of hard momentum. As a consequence,
the approximation $l_0,l\ll T$ cannot be used in this case.
Moreover, since our argument to neglect the diagram of 
Fig.~\ref{figsoft}-(c) was essentially based on the
possibility to neglect a Fermi-Dirac statistical weight in front
of a Bose-Einstein one when $l_0\ll T$, it is now impossible to
apply it. Therefore, this diagram contributes now at the same order.
\end{itemize}

Extrapolations of the previous case (despite its incompleteness)
based on Eqs.~(\ref{bigQ}),
as well as preliminary estimations of the order of magnitude
for the production rate of a static photon ($Q^2/q_0^2=1$),
seems to indicate that the bremsstrahlung
gives a contribution $\hbox{\rm Im}\,\Pi^\mu{}_\mu
\sim e^2g^4 T^3/q_0$. Therefore, the diagram of
Fig.~\ref{figsoft}-(a), already considered by Braaten, Pisarski and
Yuan, is not the only contribution to the production of
 virtual photons.

\section{Summary}

We have studied the production of photons of energy $q_0\ll T$ by 
a plasma. For virtualities verifying $Q^2/q_0^2\ll 1$, this photon
emission is dominated by the bremsstrahlung process.
In this calculation of the photon production rate, the HTL framework
appeared to be insufficient in two respects: first of all, the
standard HTL power counting breaks down in extracting the dominant 
contributions to this rate. Indeed, the diagram we considered
is supposed to be sub-dominant in the HTL expansion, but turns
out to dominate over the diagrams considered in the HTL framework.
The reason for this failure lies in a strong enhancement of
our diagrams due to collinear sensitivity.
Secondly, once the relevant contributions
have been isolated, the HTL resummation of soft lines is insufficient 
to regularize collinear divergences. This problem is
solved by the application of a recent extension of the HTL
resummation program. The 
imaginary part of the polarization tensor we get, proportional to
the 
production rate, goes
from $\hbox{\rm Im}\,\Pi^\mu{}_\mu\sim e^2g^2T^3/q_0$ when $Q^2/q_0^2\ll g^2$ to
$\hbox{\rm Im}\,\Pi^\mu{}_\mu\sim e^2g^4T^3/q_0$ when $Q^2/q_0^2\sim 1$. At the same time,
when the ratio $Q^2/q_0^2$ increases,
the bremsstrahlung process becomes more and more sensitive to gluon
exchanges of momentum intermediate between the soft scale $gT$ and 
the hard scale $T$,
which is again an important qualitative difference with the
standard HTL calculations.

When the vir\-tua\-li\-ty $Q^2/q_0^2$ approa\-ches 
$1$, the en\-han\-ce\-ment disappears,
but the bremss\-tra\-hlung process remains of an order comparable
to the contributions already calculated by Braaten, Pisarski and Yuan.
This may be interpreted as follows: the rate obtained within the HTL
expansion by BPY is unexpectedly subleading due to
a partial cancelation when taking the trace $\Pi^\mu{}_\mu$ of
the polarization tensor $\Pi^\mu{}_\nu$.
 Therefore, other terms subleading in
the HTL power counting rules may contribute as well.

\section{Acknowledgments}
One of us (EP) thanks F.~Flechsig, A.~Rebhan, and H.~Schulz for
valuable discussions. 
The work of PA and FG
is supported in part by the EEC program
``Human Capital and Mobility", Network
``Physics at High Energy Colliders",
contract CHRX-CT93-0357 (DG 12 COMA).
The work of RK and EP is supported by
the Natural Sciences and Engineering Research Council
of Canada. We also acknowledge support by NATO under grant CRG. 930739.

\section{Appendix}

To discuss the asymptotic behavior of the integrals $J_{_{T,L}}$ defined in 
Eq.~(\ref{eqjtl}),
it is convenient to split 
the integral over $w$ in $J_{_{T,L}}$ in order to obtain:
\begin{eqnarray}
&\displaystyle{J_{_{T,L}}= \int\limits_{0}^{1}\;
\frac{dx}{x}\;\widetilde{I}_{_{T,L}}(x)}\;&\left[
\int\limits_{0}^{1}\;dw\;
\frac{\sqrt{w/(w+4)}
\hbox{\rm tanh}{}^{-1}\sqrt{w/(w+4)}}{(w+\widetilde{R}_{_{T,L}}(x))^2+
(\widetilde{I}_{_{T,L}}(x))^2}\right.\nonumber\\
&&+\left.
\int\limits_{0}^{1}\;dw\;
\frac{\sqrt{1/(1+4w)}
\hbox{\rm tanh}{}^{-1}\sqrt{1/(1+4w)}}
{(1+w\widetilde{R}_{_{T,L}}(x))^2+
(w\widetilde{I}_{_{T,L}}(x))^2}\right].
\end{eqnarray} 
In each of the two terms, we will use the small $w$
approximation:
\begin{eqnarray}
&&
{\sqrt{w/(w+4)}
\hbox{\rm tanh}{}^{-1}\sqrt{w/(w+4)}}\,\build{w\ll 1}\over{\approx}\,
{{w}\over{4}}
\nonumber\\
&&
{\sqrt{1/(1+4w)}
\hbox{\rm tanh}{}^{-1}\sqrt{1/(1+4w)}}\,\build{w\ll 1}\over{\approx}\,
{{\ln(1/w)}\over 2}.
\end{eqnarray}
Moreover, since the dominant behavior of the integrals $J_{_{T,L}}$ is
controlled by the small $x$ region, we will use  
the following approximations 
\begin{eqnarray}
&&
\widetilde{R}_{_{L}}(x)\,\build{x\ll1}\over{\approx}\,
{{3m^2_{\hbox{\cmr g}}}\over{M^2_\infty}}
\qquad\qquad 
\widetilde{I}_{_{L}}(x)\,\build{x\ll1}\over{\approx}\,
-{{3m^2_{\hbox{\cmr g}}x}\over{2M^2_\infty}}
\nonumber\\
&&
\widetilde{R}_{_{T}}(x)\,\build{x\ll1}\over{\approx}\,
{{3m^2_{\hbox{\cmr g}}x^2}\over{2M^2_\infty}}
\qquad\qquad
\widetilde{I}_{_{T}}(x)\,\build{x\ll1}\over{\approx}\,
{{3m^2_{\hbox{\cmr g}}x}\over{4M^2_\infty}}.
\label{smallx}
\end{eqnarray}
The only supplementary ingredient we need is to recall that the two 
functions $\widetilde{R}_{_{T,L}}(x)$ and $\widetilde{I}_{_{T,L}}(x)$
are proportional to the ratio $m^2_{\hbox{\cmr g}}/M^2_\infty$, which
determines their order of magnitude, compared to $w$ which is
of order $1$.
After a bit of algebra, one obtains the Eqs.~(\ref{minfeq0}).

\par\section*{Figure Captions}
\begin{figure}[ht]
\begin{center}
\leavevmode
\leftskip -1cm
\epsfbox{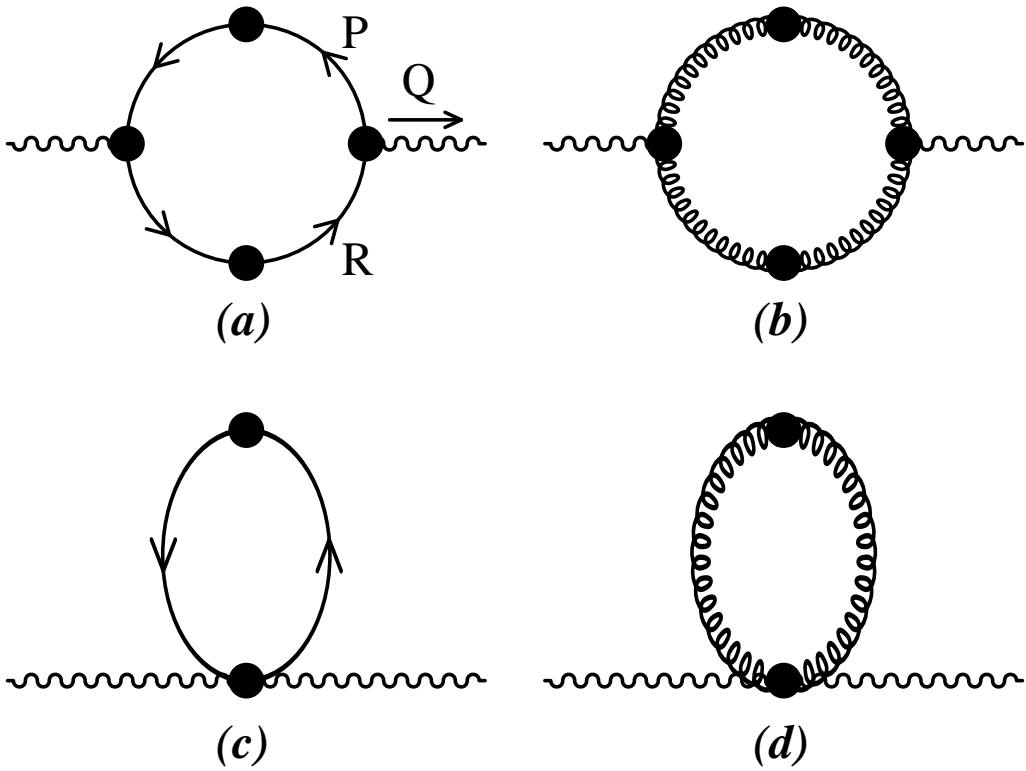}
\leftskip 0cm
\end{center}
\caption{Contributions to the soft photon production
rate with soft internal lines. The solid dots indicate
effective propagators or vertices.}
\label{figsoft}
\end{figure}

\begin{figure}[ht]
\begin{center}
\leavevmode
\epsfxsize=5 in
\vbox{\epsfbox{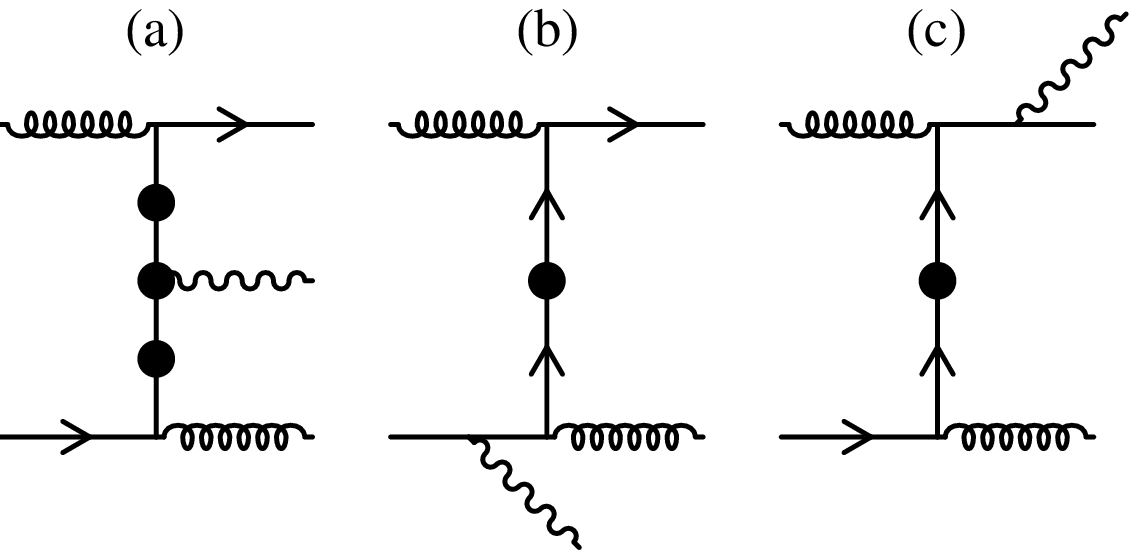}\vglue 5mm\hbox to\textwidth{\hglue 5mm
\epsfbox{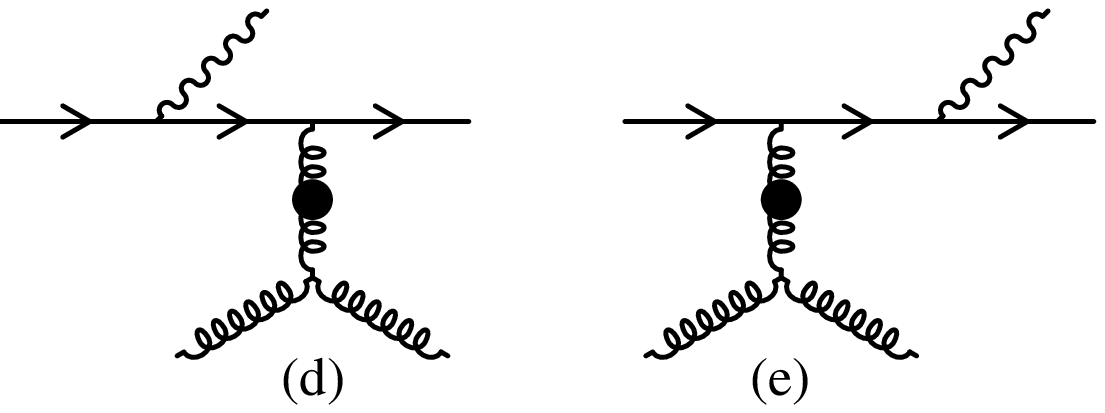}\hfill}}
\end{center}
\caption{(a),(b),(c): matrix elements contributing to the imaginary 
part
of diagrams (a) and (c) in Fig.~\ref{figsoft}; (d),(e): matrix 
elements contributing to the imaginary part
of the gluon tadpole diagram in Fig.~\ref{figsoft}.
The solid dots indicate 
effective propagators or vertices. The emitted photon has momentum
much less than $T$. The other lines carry hard momentum.}
\label{figampl}
\end{figure}

\begin{figure}[ht]
\begin{center}
\leavevmode
\epsfxsize=5 in
\vbox{\epsfbox{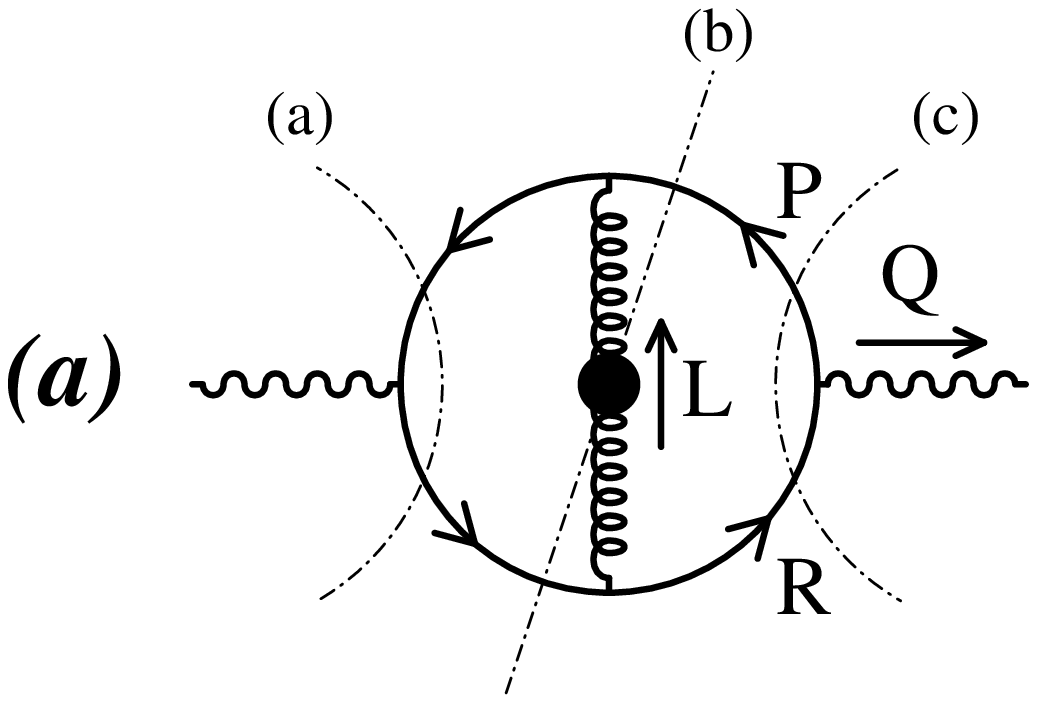}\epsfbox{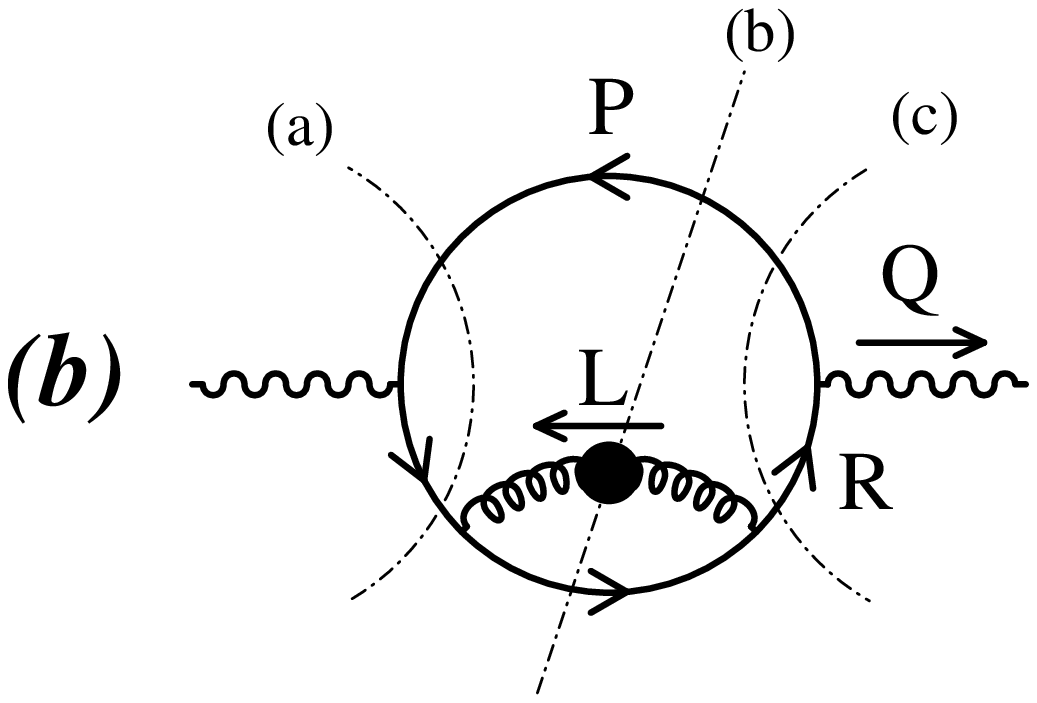}}
\end{center}
\caption{Contributions to the soft photon production
rate with hard internal fermion lines -- (a): vertex insertion;
(b): self--energy insertion. The gluon propagator is
an effective one. The fermion in the loop is hard.}
\label{fighard}
\end{figure}

\begin{figure}[ht]
\begin{center}
\leavevmode
\epsfxsize=5 in
\vbox{\epsfbox{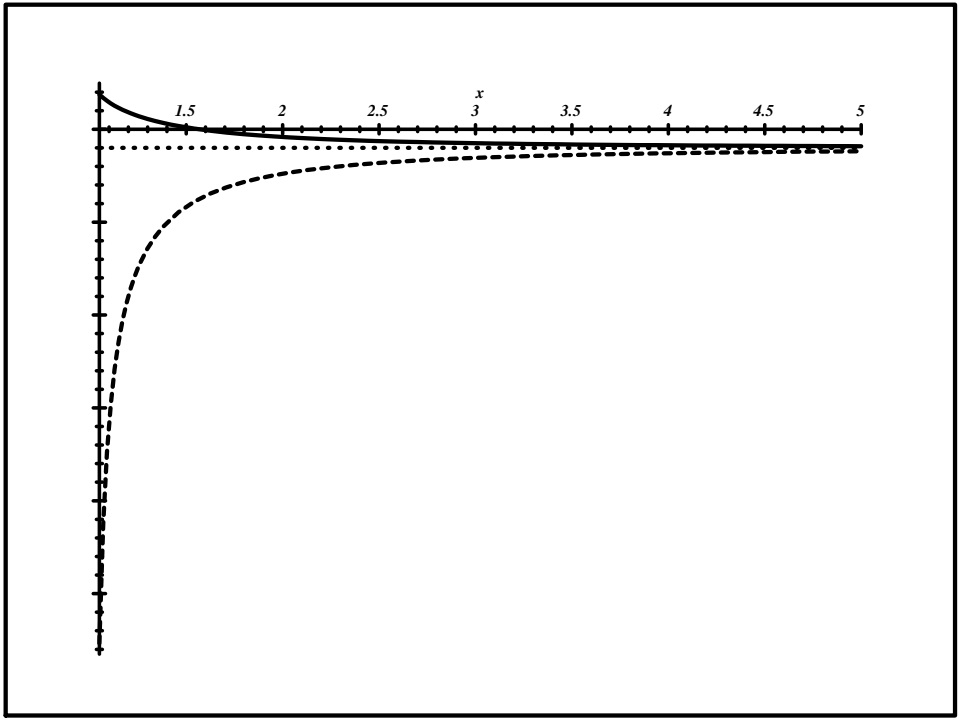}}
\end{center}
\caption{Solid line: $1-4M^2_\infty/\hbox{\rm Re}\,
\Pi_{_{T}}(x)$. Dashed line: $1-4M^2_\infty/\hbox{\rm Re}\,
\Pi_{_{L}}(x)$. Dotted line: $1-4M^2_\infty/m^2_{\cmr g}$.}
\label{figsumrules}\end{figure}

\begin{figure}[ht]
\begin{center}
\leavevmode
\epsfxsize=5 in
\vbox{\epsfbox{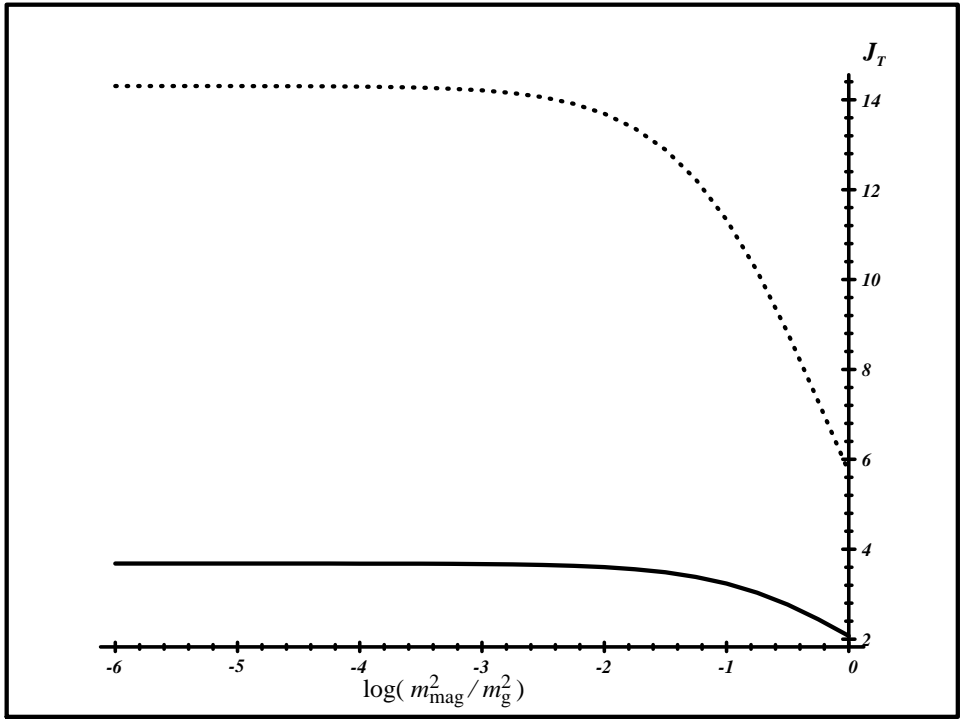}}
\end{center}
\caption{Effect of the magnetic mass on $J_{_{T}}$.
Solid line: $(M_\infty/m_{\hbox{\cmr g}})^2 = 1$.
Dotted line: $(M_\infty/m_{\hbox{\cmr g}})^2 = 0.1$.}
\label{figmag}\end{figure}

\begin{figure}[ht]
\begin{center}
\leavevmode
\epsfxsize=5 in
\vbox{\epsfbox{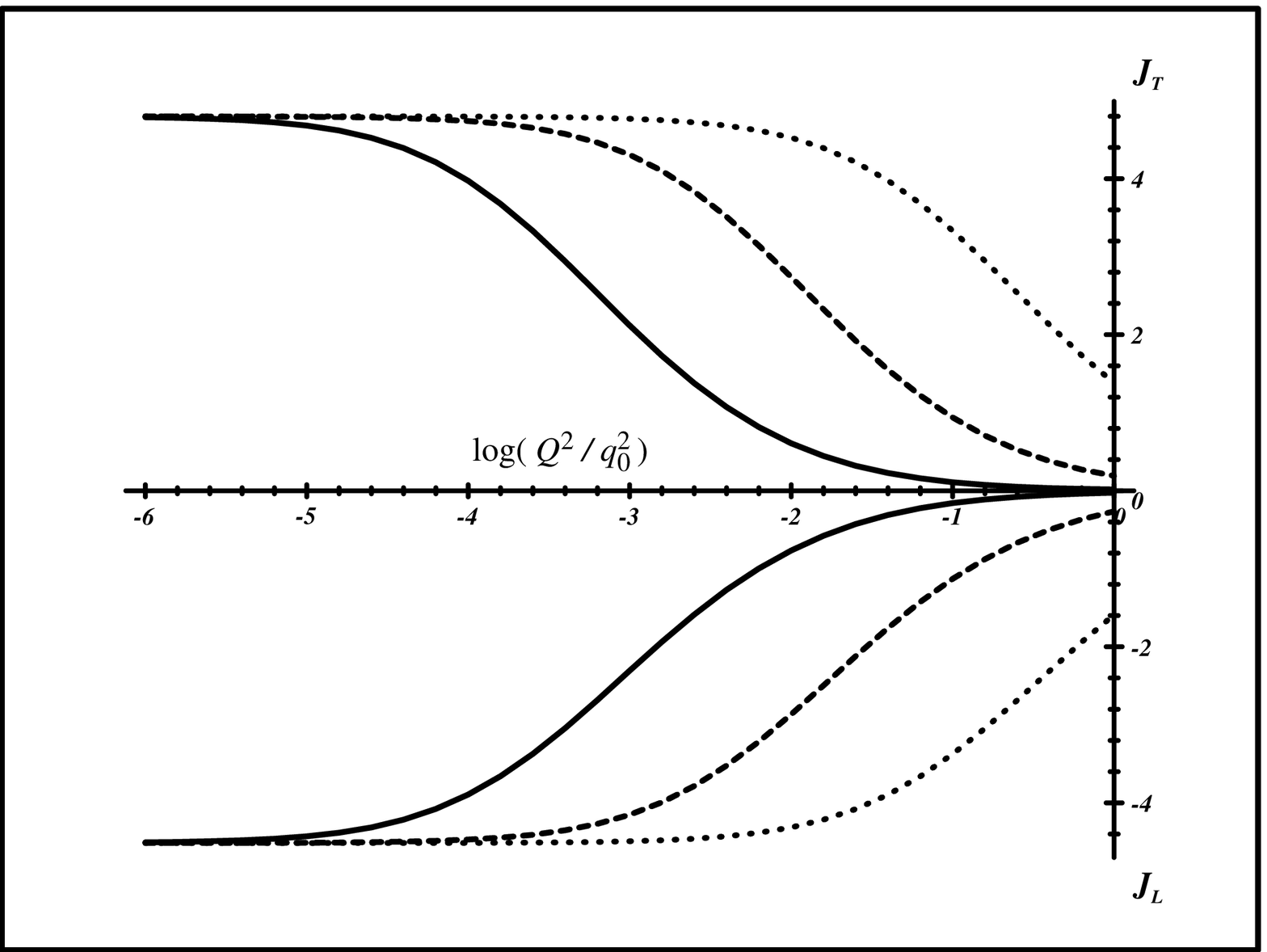}}
\end{center}
\caption{$J_{_{T,L}}$
 as a function of $\log(Q^2/q_0^2)$,
with $(m_{\hbox{\cmr g}}/M_\infty)^2$ fixed to $1.5$ (i.e. $N=3$ colors 
and 
$N_{\hbox{\cmr f}}=3$ light flavors).
Solid curves: $(m_{\hbox{\cmr g}}/T)^2= 0.005$ or $g=0.1$.
Dashed curves: $(m_{\hbox{\cmr g}}/T)^2= 0.1$ or $g=0.44$.
Dotted curves: $(m_{\hbox{\cmr g}}/T)^2= 2$ or $g=2$.
}
\label{figQ}\end{figure}

\begin{itemize}
\item[Fig.~\ref{figsoft}] Contributions to the soft photon production
rate with soft internal lines.
\item[Fig.~\ref{figampl}] a,b,c: matrix elements contributing to the imaginary part
of diagrams (a) and (c) of the previous figure; d,e: matrix elements contributing to the imaginary part
of the gluon tadpole diagram in Fig.~\ref{figsoft}.
\item[Fig.~\ref{fighard}] Contributions to the soft photon production
rate with hard internal fermion lines -- (a): vertex insertion;
(b): self--energy insertion.
\item[Fig.~\ref{figsumrules}]  $1-4M^2_\infty/\hbox{\rm Re}\,
\Pi_{_{T,L}}(x)$.
\item[Fig.~\ref{figmag}]  Effect of the magnetic mass on $J_{_{T}}$.
\item[Fig.~\ref{figQ}]  $J_{_{T,L}}$
 as a function of $\log(Q^2/q_0^2)$.
\end{itemize}
\clearpage
\clearpage
\clearpage
\clearpage
\clearpage


\begin{thebibliography}{99}

\bibitem{rp}    R.~Pisarski, Physica {\bf A158}, 146, 246 (1989);
                Phys.~Rev.~Lett.~{\bf 63}, 1129 (1989).
\bibitem{brat}  E.~Braaten and R.~D.~Pisarski,
                Nucl.~Phys.~{\bf B337}, 569 (1990); {\bf B339}, 310 (1990).
\bibitem{tay}   J.~Frenkel and J.~C.~Taylor, Nucl.~Phys.~{\bf B334}, 
                199 (1990); Z.~Phys.~{\bf C49}, 515 (1991).
\bibitem{yuan}  E.~Braaten, R.~D.~Pisarski and T.~C.~Yuan, 
                Phys.~Rev.~Lett.~{\bf 64}, 2242 (1990).
\bibitem{schiff}R.~Baier, S.~Peign\'e, and D.~Schiff, 
                Z.~Phys.~{\bf C62}, 337 (1994).
\bibitem{pat}   P.~Aurenche, T.~Becherrawy, and E.~Petitgirard,
                hep--ph/9403320 preprint (1993).
\bibitem{cleymans} J.~Cleymans, V.~V.~Goloviznin and K.~Redlich, 
                Phys.~Rev.~{\bf D47}, 989 (1993).
\bibitem{cleyman1}J.~Cleymans, V.~V.~Goloviznin and K.~Redlich, 
                Z.~Phys.~{\bf C59}, 495 (1993).
\bibitem{goloviz}V.~V.~Goloviznin and K.~Redlich, 
                Phys.~Lett.~{\bf B319}, 520 (1993).
\bibitem{lanpom}L.~D.~Landau and I.~Ya.~Pomeranchuk, Dokl.~Akad.~Nauk 
                {\bf 92}, 535 (1953); {\bf 92}, 735 (1953); \\
                A.B.~Migdal, Phys.~Rev.~{\bf 103}, 1811 (1956).
\bibitem{quack} E.~Quack and P.~A.~Henning, Phys. Rev. Lett. {\bf 75}, 2811 (1995);
hep--ph/9508201 (to appear in Phys. Rev. {\bf D}).
\bibitem{knoll} J.~Knoll and D.~N.~Voskresensky, Phys.~Lett.~{\bf B351}, 43 (1995);
                GSI-Preprint 95-63, hep--ph/9510417.
\bibitem{weldon}A.~Weldon, Phys.~Rev.~{\bf 49}, 1579 (1994).
\bibitem{gupta} S.~Gupta, D.~Indumathi, P.~Mathews and V. Ravindran, 
                Nucl.~Phys.~{\bf B4580}, 189 (1996).
\bibitem{letter}P.~Aurenche, F.~Gelis, R.~Kobes and E.~Petitgirard,
                hep--ph/9604398 preprint (to be published in Phys. Rev. {\bf D}). 
\bibitem{niega} A.~Niegawa, Mod. Phys. Lett.~{\bf A9}, 355 (1995).
\bibitem{rebhan}F.~Flechsig and A.~K.~Rebhan, 
                Nucl.~Phys.~{\bf B464}, 279 (1996).
\bibitem{rebha1}U.~Krammer, A.~K.~Rebhan and H. Schulz, Phys.~Rev.~{\bf D52}, 
                2994 (1995).
\bibitem{rebha2}A.~K.~Rebhan, Nucl.~Phys.~{\bf B430}, 319 (1994) 
\bibitem{gale}  C.~Gale and J.I.~Kapusta, Nucl.~Phys.~{\bf B357}, 65 (1991).
\bibitem{ra}    P.~Aurenche and T.~Becherrawy, Nucl.~Phys.~{\bf B379}, 259 (1992).
\bibitem{van}   C.~M.~A.~van Eijck and Ch.~G.~van Weert, Phys.~Lett. 
                {\bf B278}, 305 (1992).
\bibitem{van2}  C.~M.~A. van~Eijck, Ch.~G.~van Weert, and R.~Kobes, 
                Phys.~Rev.~{\bf D50}, 4097 (1994).
\bibitem{fett}  A.~Fetter and J.~Walecka, Quantum Theory of Many Particle
                Systems, Mc Graw and Hill.
\bibitem{niem}  A.~Niemi and G.W.~Semenoff, Ann. Phys. {\bf152} (1984) 105;
                Nucl. Phys. {\bf B230} (1984) 181.
\bibitem{kobe}  R.~Kobes and G.W.~Semenoff, Nucl. Phys. {\bf B260} (1985) 714;
                Nucl. Phys. {\bf B272} (1986) 329.
\bibitem{land}  N.~P.~Landsman and Ch.~G.~van Weert, Phys. Rep. 
                {\bf 145} (1985) 141.
\bibitem{wel}   H.~A.~Weldon, Phys.~Rev.~{\bf D26}, 1384, (1982).
\bibitem{klim}  V.~V.~Klimov, Sov.~J.~Nucl.~Phys. {\bf 33}, 934 (1981); \\
                Sov.~Phys.~JETP {\bf 55}, 199 (1982).
\bibitem{pisars}R.D.~Pisarski, Physica {\bf A158}, 146 (1989).
\bibitem{wel2}  H.~A.~Weldon, Phys.~Rev.~{\bf D26}, 2789, (1982).
\bibitem{collin}J.~Cleymans and I.~Dadic, Z.~Phys.~{\bf C42}, 133 (1989).
\bibitem{collin1}R.~Baier, B.~Pire and D.~Schiff, Phys.~Rev.~{\bf D38}, 2814 (1988); \\
                T.~Altherr, P.~Aurenche and T.~Becherrawy, Nucl.~Phys.~{\bf B315},
                436 (1989); \\ T.~Altherr and T.~Becherrawy, 
                Nucl.~Phys.~{\bf B330}, 174 (1990); \\
                Y.~Gabellini, T.~Grandou and D.~Poizat, Ann. Phys.~{\bf 202}, 436
                (1990).
\bibitem{collin2}M.~Le~Bellac and P.~Reynaud, Nucl.~Phys.~{\bf B380}, 423 (1992).
\bibitem{collin3}T.~Altherr and P.~Aurenche, Z.~Phys.~{\bf C45}, 99 (1990).
\bibitem{land-lif}L.D.~Landau, E.M.~Lifchitz, Electrodynamique quantique, Chap. X
(Cours de Physique Th\'eorique, IV), Ed.~MIR, Moscou, (1989).
\bibitem{peigne}S.~Peign\'e, Th\`ese de l'Universit\'e de Paris-Sud,
 Orsay (1995).

\end{thebibliography}
\end{document}